\documentclass[preprint, review, 12pt, sort&compress]{elsarticle}

\usepackage{amssymb}
\usepackage{amsmath}

\usepackage{numcompress}

\journal{International Journal of Mechanical Sciences}

\begin{document}

\begin{frontmatter}



\title{High- and low-entropy layers in solids behind shock and ramp compression waves\tnoteref{t1,t2}}
\tnotetext[t1]{
DOI: 10.1016/j.ijmecsci.2020.105971}
\tnotetext[t2]{
{\raisebox{-0.21\baselineskip}{\includegraphics[height=0.78\baselineskip]{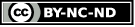}}} $\copyright$ 2020. This manuscript version is made available under the CC-BY-NC-ND 4.0 license http://creativecommons.org/licenses/by-nc-nd/4.0/.}

\author[label1,label2,label4]{K.V.~Khishchenko}\corref{cor1}\ead{konst@ihed.ras.ru}
\author[label3,label4]{A.E.~Mayer}\ead{mayer@csu.ru}
\cortext[cor1]{Corresponding author. Address for correspodence: Joint Institute for High Temperatures RAS, Izhorskaya 13 Bldg 2, Moscow 125412, Russia.}
\address[label1]{Joint Institute for High Temperatures RAS, Izhorskaya 13 Bldg 2, Moscow 125412, Russia}
\address[label2]{Moscow Institute of Physics and Technology, Institutskiy 9, Dolgoprudny, Moscow Region 141700, Russia}
\address[label4]{South Ural State University (National Research University), Lenin Prospect 76, Chelyabinsk 454080, Russia}
\address[label3]{Chelyabinsk State University, Bratyev Kashirinykh 129, Chelyabinsk 454001, Russia}



\begin{abstract}
Non-uniform temperature fields are analyzed, which arise in the problems of formation of the steady shock wave at impact and ramp loading of metals, exit of the steady shock wave to the free surface, and the shock wave passing through the interface between two different materials. Theoretical analysis and computations show that high-entropy (with the temperature increase) and low-entropy (with the temperature decrease) layers arise near the interfaces in the above problems of shock and ramp loading. The impact produces the high-entropy layer; while the ramp loading can result in the both high- and low-entropy layers. At the shock wave passing through the interface, the high-entropy layer is formed in the lower-impedance material and the low-entropy---in the higher-impedance one. The formation of high-entropy layer at impact is supported by molecular-dynamics simulations in addition to continuum modeling. The high- and low-entropy layers should be taken into account in simulations of shock-wave processes in thin targets or in other cases where surface effects are important.
\end{abstract}


\end{frontmatter}



\section{Introduction}
\label{sec1}
Dynamic loading of thin metallic foils is under increasing attention \cite{Gupta-Winey-Trivedi-LaLone-Smith-Eggert-Collins-JAP-2009, Winey-LaLone-Trivedi-Gupta-JAP-2009, Neff-Martinez-Plechaty-Stein-Presura-HEDP-2010, Ashitkov-Agranat-Kanel-Komarov-Fortov-2010-eng, Whitley-McGrane-Eakins-Bolme-Moore-Bingert-JAP-2011, Smith-Eggert-Rudd-Swift-Bolme-Collins-JAP-2011, Abrosimov-Bazhulin-Voronov-Geraskin-Krasyuk-Pashinin-Semenov-Stuchebryukhov-Khishchenko-Fortov-QE-2013-eng, Inogamov-Zhakhovsky-Petrov-Khokhlov-Ashitkov-Khishchenko-Migdal-Ilnitsky-Emirov-Komarov-Shepelev-Miller-Oleynik-Agranat-Andriyash-Anisimov-Fortov-CPP-2013, Belkov-Derkach-Garanin-Mitrofanov-Voronich-Fortov-Levashov-Minakov-JAP-2014, Ashitkov-Komarov-Struleva-Agranat-Kanel-JETPLett-2015, Krasyuk-Pashinin-Semenov-Khishchenko-Fortov-2016, Zuanetti-McGrane-Bolme-Prakash-JAP-2018} because it is the way to investigate the matter properties at extremely high strain rates. The high-rate deformation is not only a scientific issue because it takes place in novel surface treatment techniques, such as laser pinning~\cite{Wang-Wang-Xu-Gao-IJMS-2016, Hu-Xie-Wu-Yao-IJMS-2019, Angulo-Cordovilla-Garcia-Beltran-Smyth-Langer-Fitzpatrick-Ocana-IJMS-2019}, as well as in protective applications~\cite{Cao-Wang-Zhou-Yang-Wang-Pang-Chi-Su-IJMS-2019}. Plate impact is used for foils perhaps no thinner than 20~$\mu$m \cite{Gupta-Winey-Trivedi-LaLone-Smith-Eggert-Collins-JAP-2009, Winey-LaLone-Trivedi-Gupta-JAP-2009, Neff-Martinez-Plechaty-Stein-Presura-HEDP-2010}, while the ultra-short intensive laser irradiation \cite{Ashitkov-Agranat-Kanel-Komarov-Fortov-2010-eng, Whitley-McGrane-Eakins-Bolme-Moore-Bingert-JAP-2011, Smith-Eggert-Rudd-Swift-Bolme-Collins-JAP-2011, Abrosimov-Bazhulin-Voronov-Geraskin-Krasyuk-Pashinin-Semenov-Stuchebryukhov-Khishchenko-Fortov-QE-2013-eng, Inogamov-Zhakhovsky-Petrov-Khokhlov-Ashitkov-Khishchenko-Migdal-Ilnitsky-Emirov-Komarov-Shepelev-Miller-Oleynik-Agranat-Andriyash-Anisimov-Fortov-CPP-2013, Belkov-Derkach-Garanin-Mitrofanov-Voronich-Fortov-Levashov-Minakov-JAP-2014, Ashitkov-Komarov-Struleva-Agranat-Kanel-JETPLett-2015, Krasyuk-Pashinin-Semenov-Khishchenko-Fortov-2016, Wang-Wang-Xu-Gao-IJMS-2016, Zuanetti-McGrane-Bolme-Prakash-JAP-2018, Angulo-Cordovilla-Garcia-Beltran-Smyth-Langer-Fitzpatrick-Ocana-IJMS-2019} can be used for foils of several microns in thickness or even thinner; the strain rate exceeds $10^9$~s$^{-1}$ in the latter case. Dynamic shear strength and spall strength tend to the theoretical limits at such extreme loading conditions \cite{Kanel-Fortov-Razorenov-2007-eng, Ashitkov-Agranat-Kanel-Komarov-Fortov-2010-eng, Ashitkov-Komarov-Agranat-Kanel-Fortov-2013-eng, Abrosimov-Bazhulin-Bolshakov-Konov-Krasyuk-Pashinin-Ralchenko-Semenov-Sovyk-Stuchebryukhov-Fortov-Khishchenko-Khomich-QE-2014-eng, Ashitkov-Komarov-Struleva-Agranat-Kanel-JETPLett-2015, Krasyuk-Pashinin-Semenov-Khishchenko-Fortov-2016}; these strengths are determined by inertness of development of the defects subsystems---dislocations and voids correspondingly \cite{Preston-Tonks-Wallace-JAP-2003, Kanel-Fortov-Razorenov-2007-eng, Mayer-Khishchenko-Levashov-Mayer-JAP-2013, Luscher-Addessio-Cawkwell-Ramos-2017, Cao-Wang-Zhou-Yang-Wang-Pang-Chi-Su-IJMS-2019}.

Compression waves formed under the laser irradiation in thin foils are treated as ramp waves \cite{Whitley-McGrane-Eakins-Bolme-Moore-Bingert-JAP-2011, Smith-Eggert-Rudd-Swift-Bolme-Collins-JAP-2011} with a gradual increase of pressure. Such a wave should eventually breaks to the shock wave during its propagation into the bulk of target \cite{Demaske-Zhakhovsky-Inogamov-Oleynik-PRB-2010, Demaske-Zhakhovsky-Inogamov-Oleynik-PRB-2013, Inogamov-Zhakhovsky-Petrov-Khokhlov-Ashitkov-Khishchenko-Migdal-Ilnitsky-Emirov-Komarov-Shepelev-Miller-Oleynik-Agranat-Andriyash-Anisimov-Fortov-CPP-2013}, but it does not happen if the target is thin enough. The ramp waves can be created by other means as well, for example, using the magnetic pressure \cite{Hayes-Hall-Asay-Knudson-JAP-2004, Seagle-Davis-Martin-Hanshaw-APL-2013}. If the target is relatively thick and the pressure rise time is high, then the ramp-wave loading becomes nearly isentropic compression of substance because there is no discontinuity inherent to shock waves \cite{Hayes-Hall-Asay-Knudson-JAP-2004}. Shortening of the pressure rise time increases the entropy production in the ramp wave \cite{Ding-2006, Ding-Asay-JAP-2007, Ding-Asay-INTPLA-2009, Seagle-Davis-Martin-Hanshaw-APL-2013} and makes compression closer to that in a shock wave of the same intensity.

On the other hand, the shock wave is a discontinuity in analytical solution only for an ideal medium in absence of any dissipation. Shock waves in solids always have a finite rise time \cite{Johnson-Barker-JAP-1969, Swegle-Grady-JAP-1985, Kanel-JMPS-1998, Kanel-Razorenov-Baumung-Singer-JAP-2001}, that means a finite thickness of the wave front, which is determined by the dissipative processes, plasticity first of all \cite{Johnson-Barker-JAP-1969}. 
The differences from the ramp wave are as follows: the unique dependence of the entropy jump upon the pressure or velocity jump, which is determined by the Hugoniot relation \cite{Zeldovich-Raizer-1967}; uncontrolled form of the shock front, which is often supposed to be a steady one \cite{Swegle-Grady-JAP-1985}. The front thickness and strain rate of the steady shock wave are determined by the entropy jump and dissipative properties of the material---viscosity in fluids and plasticity in solids \cite{Sakharov-Zaidel-Mineev-Oleinik-1965-eng, Johnson-Barker-JAP-1969, Swegle-Grady-JAP-1985}.

An unsteady compression precedes the shock wave formation at both the plate impact and ramp loading. From the mechanical point of view, the plate impact is a kind of problem of the initial discontinuity break, rather the discontinuity of velocity field. Therefore, an initial strain rate (the particle velocity gradient) is mathematically infinite, while physically restricted by the atom size, like in the molecular dynamics (MD) simulations of the impact \cite{Bringa-Rosolankova-Rudd-Remington-Wark-Duchaineau-Kalantar-Hawreliak-Belak-2006, Belashchenko-HT-2013, Branicio-Nakano-Kalia-Vashishta-INTPLA-2013, Wang-Xiao-Deng-Zhu-Hu-INTPLA-2014}, or roughness and inclination of the colliding metal surfaces. In any case, the initial strain rate at collision is much higher than the strain rate at the steady shock. Thus, the shock wave formation at plate impact consists of gradual increase of the compression wave thickness and decrease of the corresponding strain rate with its motion inside the material. Similar to the ramp waves \cite{Ding-2006, Ding-Asay-JAP-2007, Ding-Asay-INTPLA-2009}, the strain rate decrease should be accompanied by the decrease of the entropy production and temperature rise behind the compression wave. Far from the impact surface, the entropy increment tends to the jump determined by the Hugoniot relation. Near the impact surface, the entropy increment should be higher and a high-entropy layer should be formed here. Formation of the shock wave at ramp loading is an opposite process with the gradual decrease of the compression wave thickness down to the shock front thickness; therefore a low-entropy layer should be formed in this case near the loaded surface.

The transition regions and corresponding high- and low-entropy layers can be negligibly small in comparison with the target thickness in the case of thick target. Meanwhile, those are essential for thin foils, such as used in experiments \cite{Gupta-Winey-Trivedi-LaLone-Smith-Eggert-Collins-JAP-2009, Winey-LaLone-Trivedi-Gupta-JAP-2009, Whitley-McGrane-Eakins-Bolme-Moore-Bingert-JAP-2011, Neff-Martinez-Plechaty-Stein-Presura-HEDP-2010, Ashitkov-Agranat-Kanel-Komarov-Fortov-2010-eng, Smith-Eggert-Rudd-Swift-Bolme-Collins-JAP-2011, Abrosimov-Bazhulin-Voronov-Geraskin-Krasyuk-Pashinin-Semenov-Stuchebryukhov-Khishchenko-Fortov-QE-2013-eng, Abrosimov-Bazhulin-Bolshakov-Konov-Krasyuk-Pashinin-Ralchenko-Semenov-Sovyk-Stuchebryukhov-Fortov-Khishchenko-Khomich-QE-2014-eng, Inogamov-Zhakhovsky-Petrov-Khokhlov-Ashitkov-Khishchenko-Migdal-Ilnitsky-Emirov-Komarov-Shepelev-Miller-Oleynik-Agranat-Andriyash-Anisimov-Fortov-CPP-2013, Ashitkov-Komarov-Struleva-Agranat-Kanel-JETPLett-2015, Krasyuk-Pashinin-Semenov-Khishchenko-Fortov-2016}. Therefore, formation of the high- and low-entropy layers should be investigated for adequate analysis of the experimental results. Moreover, the shock wave in solids may have an elastic precursor \cite{Kanel-JMPS-1998, Kanel-Fortov-Razorenov-2007-eng, Gupta-Winey-Trivedi-LaLone-Smith-Eggert-Collins-JAP-2009, Winey-LaLone-Trivedi-Gupta-JAP-2009, Ashitkov-Agranat-Kanel-Komarov-Fortov-2010-eng, Whitley-McGrane-Eakins-Bolme-Moore-Bingert-JAP-2011, Smith-Eggert-Rudd-Swift-Bolme-Collins-JAP-2011, Mayer-Khishchenko-Levashov-Mayer-JAP-2013, Inogamov-Zhakhovsky-Petrov-Khokhlov-Ashitkov-Khishchenko-Migdal-Ilnitsky-Emirov-Komarov-Shepelev-Miller-Oleynik-Agranat-Andriyash-Anisimov-Fortov-CPP-2013}, which decays with the distance \cite{Johnson-Barker-JAP-1969, Kanel-Razorenov-Baumung-Singer-JAP-2001}, that makes the shock wave unsteady in general. This feature probably can extend the high- and low-entropy layers in depth. 

In this paper, the high- and low-entropy layers in metals near the loaded surface are studied numerically using the dislocation plasticity model~\cite{Krasnikov-Mayer-Yalovets-IJP-2011, Mayer-Khishchenko-Levashov-Mayer-JAP-2013}. The considered surface layers are qualitatively similar to the so-called heating errors (or entropy traces, as in Russian terminology) produced by an artificial viscosity \cite{Rozhdestvensky-Yanenko-1978, Swegle-Grady-JAP-1985, Charakhchyan-CMMP-2000, Vaziev-Gadzhiev-Kuzmin-Skovpen-AIPCP-2006, Ding-2006, Ding-Asay-JAP-2007, Ding-Asay-INTPLA-2009}. The artificially-introduced viscosity or its equivalent causes a dissipation meant to ensure the stable numerical solution \cite[e.g.,][]{Lloyd-Clayton-Becker-McDowell-INTPLA-2014}. As a side effect, that leads to non-uniform distribution of entropy and temperature near interfaces after the shock wave propagation. These non-uniformities are absent in an analytical solution for ideal hydrodynamics, therefore, they are treated as shortcomings of the obtained numerical solution. Non-physical nature of such artificial entropy traces is confirmed by their strong dependence on the numerical mesh size. As opposed to the heating error, here we communicate about the physical effect caused by real dissipation processes. To exclude the mentioned numerical effect, we completely exclude the artificial viscosity and take into account only the physical, mesh-independent, viscosity and plasticity. The mesh-independent stabilizing viscous terms were previously used in~\cite{Swegle-Grady-JAP-1985, Ding-2006, Ding-Asay-JAP-2007, Ding-Asay-INTPLA-2009}. The resulting numerical solution is stable if the computational grid is fine enough, that means, if the compression wave front is resolved properly \cite{Swegle-Grady-JAP-1985}.

The paper structure is as follows. Basic relations of the entropy rise in fluids and elastic-plastic solids are considered in Section~\ref{sec2}. A mathematical model we used is described in Section~\ref{sec3}. Details of a numerical implementation of the model are presented in Section~\ref{sec4}. Numerical stability and high-entropy layers in shock-loaded fluids are analyzed in Section~\ref{sec5}. High-entropy layers in solid metals behind shock waves are investigated in Section~\ref{sec6}. Confirmation of high-entropy layers from MD simulations is considered in Section~\ref{sec7}. High- and low-entropy layers at ramp loading are studied in Section~\ref{sec8}. Interaction of the shock waves with interfaces is regarded in Section~\ref{sec9}. Finally, our conclusions are summarized in Section~\ref{sec10}.

\section{Entropy rise in fluids and solids}
\label{sec2}
As is well known, the entropy remains constant only in the equilibrium adiabatic processes, which should be very slow \cite{Bazarov-1964}. Non-equilibrium leads to the entropy rise: the higher strain rate, the greater degree of non-equilibrium, the higher entropy rise. Different dissipative processes are the particular mechanisms of such entropy rise. Here we consider two different dissipation mechanisms. Viscosity provides dissipation due to molecular pulse transfer between adjoining layers of substance. Plasticity provides dissipation due to relaxation of shear stresses by means of dislocation motion or other plasticity mechanisms.

Viscosity is the most obvious mechanism of dissipation in fluids. At uniaxial compression or tension of a fluid, the strain rate equals to $\dot{\varepsilon} = \partial \upsilon / \partial z$, where $\upsilon$ is the particle velocity and $z$ is the coordinate. Viscous stress is equal to \cite{Landau-Lifshitz-VI-1987-eng}
\begin{equation}
\sigma^\prime_{zz} = \Bigg(\frac{4\eta}{3}+\zeta\Bigg) \frac{\partial \upsilon}{\partial z} = \Bigg(\frac{4\eta}{3}+\zeta\Bigg) \dot{\varepsilon},
\end{equation}
where $\eta$ is the shear viscosity coefficient, $\zeta$ is the bulk viscosity coefficient. The dissipating power per unit volume is equal to 
\begin{equation}
\sigma^\prime_{zz} \frac{\partial \upsilon}{\partial z} = \Bigg(\frac{4\eta}{3}+\zeta\Bigg) \dot{\varepsilon}^2.
\end{equation}
This power leads to the heat release and the entropy rise with the rate
\begin{equation}
\frac{\mathrm{d} s}{\mathrm{d} t} = \Bigg(\frac{4\eta}{3}+\zeta\Bigg) \frac{\dot{\varepsilon}^2}{\rho T},
\label{eq1}
\end{equation}
where $\rho$ is the substance density, $T$ is the temperature, $s$ is the specific entropy. Rate of density change is equal to $\mathrm{d} \rho / \mathrm{d} t = -\rho \dot{\varepsilon}$, from which we obtain the following derivative:
\begin{equation}
\frac{\mathrm{d} s}{\mathrm{d} \rho} = -\Bigg(\frac{4\eta}{3}+\zeta\Bigg) \frac{\dot{\varepsilon}}{\rho^2 T}.
\label{eq2}
\end{equation}
Thus, a small density increment $\mathrm{d} \rho$ leads to the specific entropy increment $\mathrm{d} s$, which is proportional to the strain rate. So, the higher strain rate, the larger entropy rise, at the same final compression ratio.

The entropy jump on a steady shock-wave front is determined only by substance parameters on either side of the shock discontinuity that means, by the Hugoniot relation. In turn, the shock-wave thickness and the strain rate on its front are determined by the entropy jump and the substance coefficient of viscosity. The shock-wave profile becomes steady not instantly after impact: as was mentioned above, an initial velocity gradient (strain rate) is formally infinite on the interface between the colliding bodies. Although the gradient value should be physically restricted by various imperfections of the contacting surfaces or by the atomic structure of substance, nevertheless, the initial velocity gradient on the interface is higher than on the steady shock front. Therefore, as it follows from Eq.~(\ref{eq2}), the substance near the impact surface should be heated more than immediately behind the steady shock front. It is the physical reason of the high-entropy layer formation near the impact surface in a viscous fluid or similar matter with dissipation.

Similar situation takes place in solids, where an additional dissipation mechanism appears that is the plasticity. Experiments and simulations~\cite{Inogamov-Zhakhovsky-Petrov-Khokhlov-Ashitkov-Khishchenko-Migdal-Ilnitsky-Emirov-Komarov-Shepelev-Miller-Oleynik-Agranat-Andriyash-Anisimov-Fortov-CPP-2013} on the powerful ultra-short laser irradiation of thin metal foils show that the elastic-plastic properties of substance reveal themselves even in the shock waves with the pressure jump of several tens of gigapascals, at least at small target thicknesses. The high-entropy layer structure can be essential for such thin targets.

The dissipating power due to the plastic deformation per unit volume is determined by the product $S_{ik}\dot{w}_{ik}$, where $S_{ik}$ are the components of the stress deviator, $w_{ik}$ are the components of the plastic strain tensor and $\dot{w}_{ik}$ are their time derivatives; the Einstein summation convention is used. This product takes the following form at the one-dimensional deformation, which is typical for plate impact or ramp loading:
\begin{equation}
S_{ik}\dot{w}_{ik} = \frac{3}{2} S_{zz}\dot{w}_{zz},
\end{equation}
where the relations $S_{xx} = S_{yy} = -S_{zz}/2$ and $w_{xx} = w_{yy} = -w_{zz}/2$ are used, then the entropy rise rate is
\begin{equation}
\frac{\mathrm{d} s}{\mathrm{d} t} = \frac{3}{2\rho} S_{zz}\dot{w}_{zz}.
\label{eq3}
\end{equation}
The plastic strain rate $\dot{w}_{zz}$ is proportional to the dislocations velocity according to the well-known Orowan equation, while the dislocation velocity is approximately proportional to the shear stresses in the over-barrier gliding mode \cite{Kosevich-UFN-1965-eng, Merzhievsky-Tyagelsky-1988, Suzuki-Takeuchi-Yoshinaga-1991}. Therefore, one can write $\dot{w}_{zz} = K S_{zz}$, where $K$ is a coefficient of proportionality depending on the dislocations subsystem state, and, inversely, $S_{zz} = K^{-1} \dot{w}_{zz}$. This proportionality is disturbed at high values of the dislocation velocity, because the latter is bounded above by the transverse velocity of sound \cite{Merzhievsky-Tyagelsky-1988, Dudorov-Mayer-2011}, and at low values of shear stress close to the yield strength, when the thermo-fluctuation overcoming of the Peierls relief takes place with an exponential dependence on stress. The later is especially important for bcc metals with high level of the Peierls barrier. On the other hand, the fruitful application of the plasticity model with only the over-barrier dislocation gliding for fcc, bcc and hcp metals~\cite{Krasnikov-Mayer-Yalovets-IJP-2011} indicates that this mode of dislocation gliding is predominant under the shock loading for all these types of crystal lattice. The coefficient $K$ includes information about dislocation mobility and orientation of the loading direction to the dislocation slip systems. The plastic strain should completely balance the external deformation and prevent the shear stress rise at the steady plastic flow. Then one can write $\dot{w}_{zz} = 2\dot{\varepsilon}/3$, in particular, basing on expressions~\cite{Mayer-Khishchenko-Levashov-Mayer-JAP-2013}. In this case, the entropy rise is linearly dependent on the strain rate:
\begin{equation}
\frac{\mathrm{d} s}{\mathrm{d} \rho} = -\frac{2}{3K}\frac{\dot{\varepsilon}}{\rho^2 T},
\label{eq4}
\end{equation}
that is similar to Eq.~(\ref{eq2}).

Approximation of the ideal fluid (without viscosity and plasticity) is often used at the high-velocity impact modeling. The corresponding equations system is physically incomplete, which leads to instability of its numerical solution. One should use artificial viscosity \cite{vonNeumann-Richtmyer-JAP-1950} or its various equivalents to solve this problem. Such artificial dissipation increases the transition layer thickness of a strong discontinuity up to several meshes of the numerical grid. That simulates the physical viscosity, but the simulation results depend on the mesh size (for example, the artificial shock thickness) and consequently are non-physical; although, the influence of the artificial viscosity on the shock front thickness in calculations can be negligible if the numerical grid is fine enougth \cite{Lloyd-Clayton-Becker-McDowell-INTPLA-2014}. This technique seems to be founded for the gas-dynamic processes \cite{Rozhdestvensky-Yanenko-1978}, in which the physical thickness of shock waves is typically much less than the spatial resolution of the numerical grid. Meanwhile, modeling of the shock waves propagation in the condensed matter is often performed on numerical grid with the mesh size smaller than the shock front thickness. In this case, taking into account the physical viscosity and plasticity as well as elimination of the artificial ones are preferable \cite{Swegle-Grady-JAP-1985, Ding-2006, Ding-Asay-JAP-2007, Ding-Asay-INTPLA-2009}. Particularly, it is the way to obtain the physically based structure of the low- and high-entropy layers near the impact surface.

\section{Mathematical model}
\label{sec3}
In this study, we apply the dislocation plasticity model \cite{Krasnikov-Mayer-Yalovets-IJP-2011, Mayer-Khishchenko-Levashov-Mayer-JAP-2013}, which provides a quantitative description of the experimentally observed structure of shock waves and release waves \cite{Mayer-Khishchenko-Levashov-Mayer-JAP-2013, Yao-Pei-Yu-Bai-Wu-JAP-2017, Djordjevic-Vignjevic-Kiely-Case-DeVuyst-Campbell-Hughes-IJP-2018} in elastic-plastic metals including the shape and height of elastic precursor.

We consider a uniaxial deformation of elastic-plastic material (metal), which commonly takes place in the field of interest at the shock and ramp loading experiments. In the framework of continuum mechanics approach, the main equations are the continuity equation
\begin{equation}
\frac{\mathrm{d} \rho}{\mathrm{d} t} = -\rho \frac{\partial \upsilon}{\partial z},
\label{eq5}
\end{equation}
the motion equation
\begin{equation}
\rho \frac{\mathrm{d} \upsilon}{\mathrm{d} t} = \frac{\partial}{\partial z} (-P+S_{zz}+\sigma^{\prime}_{zz}),
\label{eq6}
\end{equation}
and the equation for internal energy
\begin{equation}
\rho \frac{\mathrm{d} U}{\mathrm{d} t} = (-P+\sigma^{\prime}_{zz}) \frac{\partial \upsilon}{\partial z} + (1-\alpha)\frac{3}{2} S_{zz}\frac{\mathrm{d} w_{zz}}{\mathrm{d} t} + Q + \frac{\partial}{\partial z}\Bigg(\kappa \frac{\partial T}{\partial z}\Bigg),
\label{eq7}
\end{equation}
where $P$ is the pressure; $U$ is the internal energy; $\alpha$ is the portion of the plastically dissipated energy, which is spent on formation of new dislocations, while the rest part, $(1-\alpha)$, is transformed into the heat; $Q$ is the energy release due to annihilation of dislocations; $\kappa$ is the heat conductivity coefficient. Note that the total time derivatives $\mathrm{d}/\mathrm{d} t = \partial / \partial t + \upsilon \partial / \partial z$ are used in Eqs.~(\ref{eq5})--(\ref{eq7}), which are valid for elements moving with the substance (the Lagrangian frame of reference).

The equation of motion (\ref{eq6}) takes into account both the elastic $S_{zz}$ and viscous $\sigma^\prime_{zz}$ additions to the thermodynamically equilibrium pressure $P$. The stress deviator, $S_{zz}$, describes the elastic shear stress, which remains non-zero in the deformed state even at zero strain rate. The viscous stress, $\sigma'_{zz}$, is proportional to the strain rate. The pressure does work on the density change and pumps the corresponding part $U$ of the total internal energy. The shear stresses do work on the form change and pumps another part $U_\mathrm{s}$ of the total internal energy. Some part $U_\mathrm{d}$ of the total internal energy is accumulated in structural defects (dislocations in the present case). Thus, the total internal energy is the sum $U+U_\mathrm{s}+U_\mathrm{d}$. We need to calculate explicitly the bulk part $U$ of the total energy because it is related with pressure and temperature by well-developed approach of equations of state. The part $U$ of the internal energy does not include the elastic energy of the shape change (shear deformations and stresses) and the energy of defects (dislocations). This means that $U$ corresponds to a body with the same density and temperature, but without any shear stresses and without structural defects. The second term in the right-hand part of Eq.~(\ref{eq7}) takes into account the heating at plastic deformation, similar to Eq.~(\ref{eq3}). At the same time, the elastic addition $S_{zz}$ is not included in the first term of the right-hand part of Eq.~(\ref{eq7}); otherwise that should lead to overestimation of the plastic heating. Summand $S_{zz} (\partial \upsilon / \partial z)$ pumps the elastic energy of the shape change at first, then the main part of this energy is transforming into the internal energy $U$ in the course of plastic deformation including annihilation of dislocations. This sequence of processes is reflected in Eq.~(\ref{eq7}). The last term in the right-hand part of Eq.~(\ref{eq7}) accounts for the heat conduction. So we use a more accurate energy balance than in the base model~\cite{Krasnikov-Mayer-Yalovets-IJP-2011}.

Viscous stress is determined by a simple relation \cite{Landau-Lifshitz-VI-1987-eng}:
\begin{equation}
\sigma^\prime_{zz} = \eta^* \frac{\partial \upsilon}{\partial z},
\label{eq8}
\end{equation}
where $\eta^* = 4\eta/3+\zeta$ is the coefficient of viscosity for the considered one-dimensional deformation.

Multiphase equation-of-state model for metals~\cite{Khishchenko-LiNa-JPCS-2008} is used to determine the pressure $P=P(\rho, U)$ and the temperature $T=T(\rho, U)$ as functions of density and internal energy calculated from Eqs.~(\ref{eq5}) and (\ref{eq7}) correspondingly. In the hydrodynamic approximation one should set $S_{zz} = 0$ and $w_{zz} = 0$, and then Eqs.~(\ref{eq5})--(\ref{eq8}) together with the equation of state form a closed system. In the elastic-plastic approach, one should define additional equations for the plastic strain and stress deviator instead. To describe the elastic-plastic properties, we use the dislocation plasticity model~\cite{Krasnikov-Mayer-Yalovets-IJP-2011, Mayer-Khishchenko-Levashov-Mayer-JAP-2013} as described below.

Here we make calculations only for fcc monocrystalline metals (copper and aluminum) loaded in [100] direction. In this situation, stress and strain fields are symmetrical with respect to axial rotation, and diagonal elements of the stress deviator and the plastic strain tensor satisfy the relations $S_{xx} = S_{yy} = -S_{zz}/2$ and $w_{xx} = w_{yy} = -w_{zz}/2$, while all non-diagonal elements are equal to zero. These are already taken into account in Eq.~(\ref{eq7}). The stress deviator component $S_{zz}$ is calculated from the Hook's law \cite{Landau-Lifshitz-VII-1986-eng}:
\begin{equation}
S_{zz} = 2G\Bigg( \frac{2}{3} u_{zz} - w_{zz} \Bigg),
\label{eq9}
\end{equation}
where $G$ is the shear modulus; $u_{zz}$ is the component of the macroscopic strain tensor, the time derivative of $u_{zz}$ is determined by the velocity of the macroscopic motion of matter \cite{Mayer-Khishchenko-Levashov-Mayer-JAP-2013}:
\begin{equation}
\frac{\mathrm{d} u_{zz}}{\mathrm{d} t} = \frac{\partial \upsilon}{\partial z}.
\label{eq10}
\end{equation}
The shear modulus is taken linearly increasing with pressure and linearly decreasing with temperature for both considered metals:
\begin{equation}
G(T,P) = G_0+G^\prime_P P+G^\prime_T (T-T_0),
\label{eq11}
\end{equation}
where $T_0=300$~K; $G_0$ is the shear modulus at normal conditions; the values of pressure derivative $G^\prime_P$ and temperature derivative $G^\prime_T$ are taken from \cite{Guinan-Steinberg-1974} and listed in Table~\ref{table1}.

The plastic strain rate can be calculated from the generalized Orowan equation \cite{Kosevich-UFN-1965-eng}:
\begin{equation}
\frac{\mathrm{d} w_{zz}}{\mathrm{d} t} = \sum_{\beta} b^\beta_z n^\beta_z V_{\mathrm{D} \beta} \rho_{\mathrm{D} \beta},
\label{eq12}
\end{equation}
where index $\beta$ numerates the slip systems of dislocations (the groups of dislocations) \cite{Krasnikov-Mayer-Yalovets-IJP-2011, Mayer-Khishchenko-Levashov-Mayer-JAP-2013}; $b^\beta_z$ is the component of Burgers vector and $n^\beta_z$ is the component of unit vector normal to the slip plane; $V_{\mathrm{D} \beta}$ is the velocity of moving dislocations relative to substance and $\rho_{\mathrm{D} \beta}$ is the scalar density of dislocations (the total length of dislocations of the corresponding group per unit volume of substance).

\begin{table*}[t]
\caption{\label{table1}
Parameters of the model of elastic-plastic properties: the shear modulus $G_0$ at normal conditions; the shear modulus derivatives $G^\prime_P$ and $G^\prime_T$ taken from~\cite{Guinan-Steinberg-1974}; the portion of plastically dissipated energy $\alpha$, which is spent on formation of new dislocations; the yield strength constant $Y_0$; the phonon drag parameter $\theta$; the constant of interaction $A_\mathrm{I}$; the velocity of immobilization $V_\mathrm{I}$ and the minimal density $\rho_0$ taken from~\cite{Mayer-Khishchenko-Levashov-Mayer-JAP-2013}, where those were fitted to existing experimental profiles of free surface velocity at collision of metal plates; the annihilation constant $k_a$; the coefficient of viscosity $\eta^*$.
}
\begin{center}
\begin{tabular}{lccccccccccc}
\hline
Parameter & Al & Cu \\
\hline
$G_0$, GPa & 23.3 &47.7 \\
$G^\prime_P$ & 1.8 &1.35 \\
$G^\prime_T$, GPa/kK & $-17.2$ &$-18$ \\
$\alpha$ & 0.1 &0.1 \\
$Y_0$, MPa & 22 &30 \\
$\theta$, K & 430 &280 \\
$A_\mathrm{I}$ & 6 &4 \\
$V_\mathrm{I}$, m/s & 1.7 &0.7 \\
$\rho_0$, cm$^{-2}$ & $10^8$ &$10^8$ \\
$k_a$ & 7 &10 \\
$\eta^*$, Pa\,s &1 &1 \\
\hline
\end{tabular}
\end{center}
\end{table*}

\begin{table}[t]
\caption{\label{table2}
Slip systems in fcc metals according to~\cite{Hirth-Lothe-1982}.
}
\begin{center}
\begin{tabular}{cccccc}
\hline
$n_x^\beta \sqrt{3}$ & $n_y^\beta \sqrt{3}$ & $n_z^\beta \sqrt{3}$ & $b_x^\beta \sqrt{2}/b$ & $b_y^\beta \sqrt{2}/b$ & $b_z^\beta \sqrt{2}/b$ \\
\hline
$\phantom{-}1$ & $\phantom{-}1$ & 1 & $-1$ & $\phantom{-}0$ & 1 \\
& & & $\phantom{-}0$ & $-1$ & 1 \\
& & & $-1$ & $\phantom{-}1$ & 0 \\
\hline
$-1$ & $\phantom{-}1$ & 1 & $\phantom{-}1$ & $\phantom{-}0$ & 1 \\
& & & $\phantom{-}0$ & $-1$ & 1 \\
& & & $\phantom{-}1$ & $\phantom{-}1$ & 0 \\
\hline
$\phantom{-}1$ & $-1$ & 1 & $-1$ & $\phantom{-}0$ & 1 \\
& & & $\phantom{-}0$ & $\phantom{-}1$ & 1 \\
& & & $\phantom{-}1$ & $\phantom{-}1$ & 0 \\
\hline
$-1$ & $-1$ & 1 & $\phantom{-}1$ & $\phantom{-}0$ & 1 \\
& & & $\phantom{-}0$ & $\phantom{-}1$ & 1 \\
& & & $-1$ & $\phantom{-}1$ & 0 \\
\hline
\end{tabular}
\end{center}
\end{table}

The slip systems are determined by the crystal structure of specified metal; there are 12 different slip systems in the considered fcc metals \cite{Hirth-Lothe-1982}, those are listed in Table~\ref{table2}. In the case of [100] loading, the lab coordinates coincides with the crystallographic coordinates. As one can see, four slip systems make zero contribution in the plastic strain because of $b^\beta_z=0$, while the other eight slip systems make equal contribution because the relation $b^\beta_z n^\beta_z = 6^{-1/2} b$ takes place for all of them. Therefore, a simpler form of Eq.~(\ref{eq12}) can be used at such loading conditions:
\begin{equation}
\frac{\mathrm{d} w_{zz}}{\mathrm{d} t} = \frac{b}{\sqrt{6}} V_{\mathrm{D}} \rho_{\mathrm{D}},
\label{eq13}
\end{equation}
where $\rho_{\mathrm{D}}$ is the total scalar density of dislocation in all eight active slip systems; $V_{\mathrm{D}}$ is the moving dislocation velocity in the active slip systems. The modulus of Burgers vector $b$ in the compressed substance can be calculated from the density $\rho$; for fcc crystals one can obtain:
\begin{equation}
b = \frac{1}{\sqrt{2}} \Bigg(\frac{4\mu}{\rho N_{\mathrm{A}}}\Bigg)^{1/3},
\label{eq14}
\end{equation}
where $\mu$ is the molar mass; $N_\mathrm{A}$ is the Avogadro constant.

Velocity of moving dislocations is determined by the macroscopic shear stresses \cite{Kosevich-UFN-1965-eng} characterized by $S_{zz}$. In the considered case, the equation of dislocation motion \cite{Dudorov-Mayer-2011, Mayer-Khishchenko-Levashov-Mayer-JAP-2013} takes the following form:
\begin{equation}
m_0 \frac{\mathrm{d} V_\mathrm{D}}{\mathrm{d} t} = \frac{b}{2}[1 - (V_{\mathrm{D}}/c_\mathrm{t})^2]^{3/2} \Bigg[\sqrt{\frac{3}{2}}S_{zz} - Y \,\mathrm{sign}\,(S_{zz}) \Bigg] -B V_{\mathrm{D}},
\label{eq15}
\end{equation}
where $m_0=\rho b^2$ is the rest mass of dislocations per unit length; $Y$ is the static yield strength; $B$ is the drag coefficient. The motion of dislocations and the corresponding plastic deformation take place only if $|S_{zz}| > Y\sqrt{2/3}$. The first term in the right-hand part of Eq.~(\ref{eq15}) includes a quasi-relativistic multiplier \cite{Dudorov-Mayer-2011}, which reflects the fact that the dislocation velocity is restricted above by the transverse velocity of sound $c_\mathrm{t} = \sqrt{G/\rho}$.

The phonon drag is very important in the temperature range of interest \cite{Alshitz-Indenbom-UFN-1975-eng, Suzuki-Takeuchi-Yoshinaga-1991}. The following temperature dependence can be written for the drag coefficient \cite{Suzuki-Takeuchi-Yoshinaga-1991}:
\begin{equation}
B = \frac{4 \theta^2}{h^2}\Bigg(\frac{k_\mathrm{B}}{c_\mathrm{b}}\Bigg)^3 T,
\label{eq16}
\end{equation}
where $k_\mathrm{B}$ is the Boltzmann constant; $h$ is the Planck constant; $\theta$ is the parameter with the dimensionality of temperature (see Table~\ref{table1}); $c_\mathrm{b}$ is the bulk velocity of sound, which is defined from the equation of state. The yield strength is determined by the Taylor relation \cite{Suzuki-Takeuchi-Yoshinaga-1991}:
\begin{equation}
Y = Y_0+A_\mathrm{I}Gb\sqrt{\rho_\mathrm{I}},
\label{eq17}
\end{equation}
where $Y_0$ includes resistance from point obstacles and Peierls barrier and the second term accounts interaction between moving dislocations and immobilized ones; $A_\mathrm{I}$ is the constant of interaction and $\rho_\mathrm{I}$ is the total density of immobilized dislocations \cite{Mayer-Khishchenko-Levashov-Mayer-JAP-2013}. Despite the fact that the Taylor hardening is commonly treated as the function of the total dislocation density \cite{Davoudi-Vlassak-2018, Kositski-Mordehai-2019}, here we use the density of immobilized dislocations instead. We suppose that the immobilized dislocations form behind the shock front some structures, which strongly resist the motion of the rest (mobile) dislocations. This strong resistance is reflected by the high value of the interaction constant $A_\mathrm{I}$ (see Table~\ref{table1}). On this background, the resistance of the mobile dislocation is negligible, and we neglect it for the sake of simplicity. This approach with the strong resistance from immobilized dislocations allows us to get a good fit with experimental profile of unloading wave besides the shock wave~\cite{Mayer-Khishchenko-Levashov-Mayer-JAP-2013}.

The fitted in~\cite{Mayer-Khishchenko-Levashov-Mayer-JAP-2013} values of the dislocation interaction constant $A_\mathrm{I}$ (6 for Al and 4 for Cu, see Table~1) are about order of magnitude higher than that follows from the discreet dislocation dynamics \cite{Kocks-Mecking-PMS-2003, Madec-Kubin-AM-2017}. The more intensive strengthening can be explained by formation of specific dislocation structures in the shock-loaded metals, which are more resistant to the dislocation motion than that are considered in the discreet dislocation simulations. It should be mentioned that the same dislocation kinetics with the same order of magnitude of the interaction coefficient were successfully used in \cite{Yao-Pei-Yu-Bai-Wu-JAP-2017} in order to explain the dynamics of release of shock-loaded Al in experiments~\cite{Huang-Asay-JAP-2007}.

The kinetics equations~\cite{Mayer-Khishchenko-Levashov-Mayer-JAP-2013} for mobile and immobilized dislocations are as follows:
\begin{gather}
\frac{\mathrm{d} \rho_\mathrm{D}}{\mathrm{d} t} = Q_{\mathrm{D}} - Q_{\mathrm{I}} - Q_{\mathrm{Da}} +\frac{\rho_{\mathrm{D}}}{\rho}\frac{\mathrm{d} \rho}{\mathrm{d} t},
\label{eq18}
\\
\frac{\mathrm{d} \rho_\mathrm{I}}{\mathrm{d} t} = Q_{\mathrm{I}} - Q_{\mathrm{Ia}} +\frac{\rho_{\mathrm{I}}}{\rho}\frac{\mathrm{d} \rho}{\mathrm{d} t},
\label{eq19}
\end{gather}
where $Q_{\mathrm{D}}$ is the generation rate of the mobile dislocations; $Q_{\mathrm{I}}$ is the rate of immobilization; $Q_{\mathrm{Da}}$ and $Q_{\mathrm{Ia}}$ are the annihilation rates of mobile and immobilized dislocations correspondingly:
\begin{equation}
Q_{\mathrm{Da}} = k_{\mathrm{a}} b |V_{\mathrm{D}}| \rho_{\mathrm{D}} (2\rho_{\mathrm{D}}+\rho_{\mathrm{I}}),
\quad
Q_{\mathrm{Ia}} = k_{\mathrm{a}} b |V_{\mathrm{D}}| \rho_{\mathrm{D}} \rho_{\mathrm{I}},
\label{eq21}
\end{equation}
where $k_{\mathrm{a}}$ is the annihilation factor.

The last terms in Eqs.~(\ref{eq18}) and (\ref{eq19}) take into account the dislocation density change due to compression or tension of substance. The rate of generation is as follows:
\begin{equation}
Q_{\mathrm{D}} = \frac{\alpha}{\varepsilon_\mathrm{D}} \Bigg\{\frac{BV_\mathrm{D}^2}{[1-(V_\mathrm{D}/c_\mathrm{t})^2]^{3/2}}+\frac{b}{2}Y |V_\mathrm{D}| \Bigg\} \rho_{\mathrm{D}},
\label{eq22}
\end{equation}
where $\varepsilon_\mathrm{D}\approx \mathrm{8~eV}/b$ is the energy of the dislocations formation per unit length \cite{Kittel-2004}. The multiplier in curly brackets in Eq.~(\ref{eq22}) is the energy dissipation rate per unit length of dislocation; this is the sum of the work against the phonon friction and the work against the resistance force \cite{Dudorov-Mayer-2011}. The part of dissipated energy $\alpha$ spent on formation of new dislocations is an important parameter of the model; the value $\alpha=0.1$ is used (see Table~\ref{table1}), which follows from calorimetric measurements at small plastic deformations \cite{Kittel-2004}. If one neglects the kinetic energy of dislocations $\rho_\mathrm{D} m_0 V_\mathrm{D}^2/(2\rho) = \rho_\mathrm{D} V_\mathrm{D}^2 b^2/2 \approx 10$~J/kg, then the rate~(\ref{eq22}) takes a simple form:
\begin{equation}
Q_{\mathrm{D}} = \frac{\alpha}{\varepsilon_\mathrm{D}} \Bigg(\frac{3}{2}S_{zz} \frac{\mathrm{d} w_{zz}}{\mathrm{d} t} \Bigg).
\label{eq23}
\end{equation}
The rate of immobilization is as follows:
\begin{equation}
Q_{\mathrm{I}} = V_\mathrm{I} (\rho_\mathrm{D}- \rho_0)\sqrt{\rho_\mathrm{I}}.
\label{eq24}
\end{equation}

The latter equation describes the immobilization process, where parameter $\rho_0$ is the minimal dislocations density, which is necessary for their consolidation and immobilization in dislocation structures \cite{Mayer-Khishchenko-Levashov-Mayer-JAP-2013}. This expression is written from the assumption that all excess mobile dislocations will be immobilized in structures with the characteristic time $\tau_\mathrm{I}\approx r_\mathrm{I}/V_\mathrm{I}$, where $r_\mathrm{I}\approx (\rho_\mathrm{I})^{-1/2}$ is the average distance between the immobile dislocations. Parameter $V_\mathrm{I}$ means a characteristic velocity of the dislocations movement during the process of consolidation; that is determined by internal stresses.

The substance heating rate $Q$ due to annihilation of dislocations can be expressed as
\begin{equation}
Q = \varepsilon_\mathrm{D} (Q_\mathrm{Da}+ Q_\mathrm{Ia}).
\label{eq25}
\end{equation}

So, we have formulated the closed system of equations for description of elestoplastic deformation of fcc metals on the basis of dislocations dynamics and kinetics \cite{Krasnikov-Mayer-Yalovets-IJP-2011, Mayer-Khishchenko-Levashov-Mayer-JAP-2013}. Main equations are Eqs.~(\ref{eq5})--(\ref{eq10}), (\ref{eq13}), (\ref{eq15}), (\ref{eq18}) and (\ref{eq19}). The verified model parameters are listed in Table~\ref{table1}.

\section{Numerical implementation}
\label{sec4}
In the problem, one should clearly understand the difference between the physical and numerical effects; therefore, here, a large attention is devoted to the description of the numerical implementation. The four physical processes are separated to simplify the problem: (i) the motion of a substance according to Eqs.~(\ref{eq5})--(\ref{eq8}) and (\ref{eq10}) with the reduced Eq. (\ref{eq7}) without heat conduction and plastic heating; (ii) the kinetics of dislocations and plastic deformation by Eqs.~(\ref{eq13}), (\ref{eq15}), (\ref{eq18}) and (\ref{eq19}) with auxiliary relations and the plastic heating in Eq.~(\ref{eq7}); (iii) the heat conductivity from Eq.~(\ref{eq7}); (iv) determination of the thermodynamic state from the equation of state and calculation of shear stress in accordance with Eq.~(\ref{eq9}). Contributions of processes (i)--(iii) are calculated sequentially at every time step. At numerical discretization, the substance velocities and coordinates are defined in the mesh points and numerated by the integer subscripts $\{i\}$, while all other quantities, including the thermodynamic parameters and dislocation characteristics, are defined in the centers of meshes and numerated by the half-integer subscripts $\{i+1/2\}$. The time layers are numerated by the integer superscripts $(n)$. The Lagrangian frame of reference is used as it is mentioned in the previous section.

Accounting for the physical viscosity, Eq.~(\ref{eq8}), lets one to obtain the stable numerical solution for the process (i) even by using of the simple explicit Euler method. Equation of motion (\ref{eq6}) leads to the following relation:
\begin{multline}
\upsilon_{\{i\}}^{(n+1)} = \upsilon_{\{i\}}^{(n)} 
\\
+ \Big[(-P+S_{zz}+\sigma^\prime_{zz})_{\{i+1/2\}}^{(n)} -(-P+S_{zz}+\sigma^\prime_{zz})_{\{i-1/2\}}^{(n)} \Big] \frac{\Delta t}{m_{\{i\}}},
\label{eq26}
\end{multline}
where $\Delta t = t^{(n+1)}-t^{(n)}$ is the time step; 
\begin{gather}
m_{\{i\}}=\frac{m_{\{i+1/2\}}+m_{\{i-1/2\}}}{2}, 
\\
m_{\{i+1/2\}}= \rho_{\{i+1/2\}}^{(n)} V_{\{i+1/2\}}^{(n)}
\end{gather}
is the substance mass in the mesh, that is constant during the substance motion; 
\begin{equation}
V_{\{i+1/2\}}^{(n)}=z_{\{i+1\}}^{(n)}-z_{\{i\}}^{(n)}
\end{equation}
is the mesh volume. New position of the mesh points are calculated from the following expression:
\begin{equation}
z_{\{i\}}^{(n+1)}=z_{\{i\}}^{(n)}+\Big(\upsilon_{\{i\}}^{(n+1)}+\upsilon_{\{i\}}^{(n)}\Big)\frac{\Delta t}{2}.
\label{eq27}
\end{equation}
Equation for internal energy (\ref{eq7}) gives one:
\begin{equation}
\tilde{\tilde{U}}_{\{i+1/2\}}^{(n+1)} = U_{\{i+1/2\}}^{(n)} + \frac{\Delta V_{\{i+1/2\}}}{m_{\{i+1/2\}}} (-P+\sigma^\prime_{zz})_{\{i+1/2\}}^{(n)},
\label{eq28}
\end{equation}
where
\begin{equation}
\Delta V_{\{i+1/2\}}=V_{\{i+1/2\}}^{(n+1)}-V_{\{i+1/2\}}^{(n)}
\end{equation}
is the mesh volume change over the time step. Relation (\ref{eq28}) does not take into account the internal energy change due to the plastic heating and the heat conduction on the stage (i). The substance density can be calculated from the mesh volume:
\begin{equation}
\rho_{\{i+1/2\}}^{(n+1)}=\frac{m_{\{i+1/2\}}}{V_{\{i+1/2\}}^{(n+1)}}.
\label{eq29}
\end{equation}
Approximation of the viscous stress, Eq.~(\ref{eq8}), is as follows:
\begin{equation}
(\sigma^\prime_{zz})_{\{i+1/2\}}^{(n+1)}=\eta^*\frac{\upsilon_{\{i+1\}}^{(n+1)}-\upsilon_{\{i\}}^{(n+1)}}{V_{\{i+1/2\}}^{(n+1)}},
\label{eq30}
\end{equation}
and relation for the macroscopic strain, Eq.~(\ref{eq10}), gives:
\begin{equation}
(u_{zz})_{\{i+1/2\}}^{(n+1)}=(u_{zz})_{\{i+1/2\}}^{(n)}+ \frac{\Delta z_{\{i+1\}}-\Delta z_{\{i\}}}{V_{\{i+1/2\}}^{(n+1)}},
\label{eq31}
\end{equation}
where 
\begin{equation}
\Delta z_{\{i\}}=z_{\{i\}}^{(n+1)}-z_{\{i\}}^{(n)}
\end{equation}
is the mesh point displacement over the time step.

The time step for process (i) is chosen from Courant--Friedrichs--Lewy condition:
\begin{equation}
\Delta t=0.05 \min_i \frac{2\big|z_{\{i+1\}}^{(n)}-z_{\{i\}}^{(n)}\big|}{2(c_\mathrm{l})_{\{i+1/2\}}^{(n)}+\big|\upsilon_{\{i+1\}}^{(n)}\big|+\big|\upsilon_{\{i\}}^{(n)}\big|},
\label{eq32}
\end{equation}
where
$$c_\mathrm{l}=\sqrt{c_\mathrm{b}^2+\frac{4G}{3\rho}}$$
is the longitudinal sound velocity. The coefficient 0.05 in Eq.~(\ref{eq32}) is chosen so low to ensure the correct behavior of the numerical solution at the presence of phase transitions. The same time step is used at the numerical solution for stages (ii) and (iii).

The process (ii) does not include any spatial transfer and all equations are local: ranges of dislocations are supposed to be negligibly short. Equation for the dislocations velocity (\ref{eq15}) is the most problematic in the block (ii) due to the quasi-relativistic factor and low rest mass of dislocation. Therefore, the interpolating analytic solution~\cite{Dudorov-Mayer-2011} is used instead of the precise one:
\begin{equation}
(V_\mathrm{D})_{\{i+1/2\}}^{(n+1)} = (V_\mathrm{D})_{\{i+1/2\}}^{(n)} \mathrm{e}^{-\Delta t/\tau} + V_\infty \Big(1 - \mathrm{e}^{-\Delta t/\tau}\Big),
\label{eq33}
\end{equation}
where
\begin{gather}
V_\infty=\frac{1}{6\sqrt{6\chi\xi}} \Big(\xi^{2/3}-12\Big)^{3/2}(c_\mathrm{t})_{\{i+1/2\}}^{(n)}, 
\\
\chi=\Bigg(\frac{b}{2 c_\mathrm{t} B} \Bigg[\sqrt{\frac{3}{2}}S_{zz}-Y\,\mathrm{sign}\,(S_{zz}) \Bigg] \Bigg)_{\{i+1/2\}}^{(n)},
\end{gather}
$\xi=108\chi+12 \sqrt{12+81\chi^2}$, and the inverse characteristic time
\begin{equation}
\frac{1}{\tau}=\Bigg(\frac{B}{m_0}+\frac{b}{2 c_\mathrm{t} m_0} \Bigg[\sqrt{\frac{3}{2}}S_{zz}-Y\,\mathrm{sign}\,(S_{zz}) \Bigg] \Bigg)_{\{i+1/2\}}^{(n)}.
\end{equation}

Equations for scalar density of mobile (\ref{eq18}) and immobilized (\ref{eq19}) dislocations are solved by the explicit Euler method: 
\begin{gather}
(\rho_\mathrm{D})_{\{i+1/2\}}^{(n+1)} = (\rho_\mathrm{D})_{\{i+1/2\}}^{(n)} \sigma_{\{i+1/2\}}^{(n+1/2)} + (Q_\mathrm{D} - Q_\mathrm{I}- Q_\mathrm{Da})_{\{i+1/2\}}^{(n)} \Delta t,
\label{eq34}
\\
(\rho_\mathrm{I})_{\{i+1/2\}}^{(n+1)} = (\rho_\mathrm{I})_{\{i+1/2\}}^{(n)} \sigma_{\{i+1/2\}}^{(n+1/2)} + (Q_\mathrm{I} - Q_\mathrm{Ia})_{\{i+1/2\}}^{(n)} \Delta t,
\label{eq35}
\end{gather}
where the ratio
\begin{equation}
\sigma_{\{i+1/2\}}^{(n+1/2)}=\frac{V_{\{i+1/2\}}^{(n)}}{V_{\{i+1/2\}}^{(n+1)}}
\end{equation}
takes into account the substance compression or tension according to the last terms in Eqs.~(\ref{eq18}) and (\ref{eq19}). Equation for the plastic strain (\ref{eq13}) is integrated similarly:
\begin{equation}
(w_{zz})_{\{i+1/2\}}^{(n+1)} = (w_{zz})_{\{i+1/2\}}^{(n)} + (b V_\mathrm{D} \rho_\mathrm{D})_{\{i+1/2\}}^{(n)} \frac{\Delta t}{\sqrt{6}}.
\label{eq36}
\end{equation}
Correction of the internal energy accounted for the heating at plastic deformation is as follows:
\begin{multline}
\tilde{U}_{\{i+1/2\}}^{(n+1)} = \tilde{\tilde{U}}_{\{i+1/2\}}^{(n+1)} 
\\
+Q_{\{i+1/2\}}^{(n)} \Delta t+ (1-\alpha) \frac{3}{2} (S_{zz})_{\{i+1/2\}}^{(n)} (\Delta w_{zz})_{\{i+1/2\}},
\label{eq37}
\end{multline}
where
\begin{equation}
(\Delta w_{zz})_{\{i+1/2\}}=(w_{zz})_{\{i+1/2\}}^{(n+1)}-(w_{zz})_{\{i+1/2\}}^{(n)}
\end{equation}
is the plastic strain increment over the time step.

Explicit integration is also used for the heat conductivity process (iii). The internal energy from Eq.~(\ref{eq37}) is corrected and the final value is obtained as
\begin{equation}
U_{\{i+1/2\}}^{(n+1)}=\tilde{U}_{\{i+1/2\}}^{(n+1)}-\frac{q_{\{i+1\}}^{(n)}-q_{\{i\}}^{(n)}}{z_{\{i+1\}}^{(n)}-z_{\{i\}}^{(n)}}\frac{\Delta t}{\rho_{\{i+1/2\}}^{(n)}}, 
\label{eq38}
\end{equation}
where the heat flux through the mesh boundary is
\begin{equation}
q_{\{i\}}^{(n)}=-\Big(\kappa_{\{i+1/2\}}^{(n)}+\kappa_{\{i-1/2\}}^{(n)}\Big) \frac{T_{\{i+1/2\}}^{(n)}-T_{\{i-1/2\}}^{(n)}}{z_{\{i+1\}}^{(n)}-z_{\{i-1\}}^{(n)}}.
\label{eq39}
\end{equation}
Temperature dependence of the heat conductivity coefficient is taken in tabular form~\cite{Grigoriev-Meilikhov-1997-eng}.

Stage (iv) is the implementation of the constitutive relations. The equation of state~\cite{Khishchenko-LiNa-JPCS-2008} was used in tabular form of functions $P=P(\rho,T)$, $U=U(\rho,T)$ and $c_\mathrm{b}=c_\mathrm{b}(\rho,T)$~\cite{Levashov-Khishchenko-AIPCP-2007}. Use of appropriate subroutines for search and interpolation of tabulars allows determining the temperature, pressure and bulk sound velocity from the known density via Eq.~(\ref{eq29}) and internal energy via Eqs.~(\ref{eq28}), (\ref{eq37}) and (\ref{eq38}). Then the new value of shear modulus can be calculated from Eq.~(\ref{eq11}), and the new value of stress deviator is followed from the Hook law, Eq.~(\ref{eq9}):
\begin{equation}
(S_{zz})_{\{i+1/2\}}^{(n+1)}=2 G_{\{i+1/2\}}^{(n+1)} \Bigg(\frac{2}{3} u_{zz} - w_{zz}\Bigg)_{\{i+1/2\}}^{(n+1)}.
\label{eq40}
\end{equation}

\section{Numerical stability and high-entropy layers in fluids}
\label{sec5}
We solve numerically the problem, which is equivalent to symmetric collision of metal plates. One of the plates is under consideration, which moves along the $oZ$ axis with the initial velocity $-u$. Left boundary $z=0$ of the plate is at rest, while its right boundary is free. Substance deceleration at the left boundary forms a shock wave spreading to the right; the total velocity jump on the shock wave is equal to $u$. In this section, a solution is done in hydrodynamic approximation; that means a viscous fluid is considered with thermodynamic properties of the metal.

The numerical experiments have shown that solution is unstable for non-viscous fluid (at $\eta^{*}=0$): strong oscillations of parameters take place behind the shock wave front. In viscous fluid (at $\eta^{*}>0$), the viscosity stabilizes the solution and makes the shock front thicker. Dependence of the shock front thickness $h_\mathrm{SW}$ upon the coefficient of viscosity $\eta^{*}$ and the velocity jump $u$ has been numerically investigated. The shock front thickness is calculated as a distance between two points corresponding to the substance velocity change from its initial value on the magnitude of $\Delta \upsilon/u = 0.1$ and 0.9. The obtained results correspond to the relation that follows from the dimensional considerations:
\begin{equation}
h_\mathrm{SW}=A \frac{\eta^{*}}{\rho_0 u},
\label{eq41}
\end{equation}
where $\rho_0$ is the initial substance density, $A$ is dimensionless coefficient of the order of unity: $A\approx 1.3$ for copper and $A\approx 1.4$ for aluminum.

The numerical solution is stable if the shock front layer is described by no less than 3 mesh points of the computational grid: 
\begin{equation}
h_\mathrm{SW}\geqslant 3\Delta z,
\label{eq42}
\end{equation}
where $\Delta z$ is the mesh size. The condition (\ref{eq42}) can be always satisfied by choosing of the fine enough grids. Therefore, a stable solution can be obtained at arbitrary small viscosity coefficient $\eta^{*}$ if the computational grid is detailed enough to describe the shock wave structure. The solution is unstable at large values of viscosity $\eta^{*}$ if the shock front thickness $h_\mathrm{SW}$ becomes comparable with the size of the entire computational domain. The numerical experiments show that the shock front thickness tends to zero if the viscosity coefficient tends to zero, in accordance with Eq.~(\ref{eq41}).

\begin{figure}[t]
\begin{center}
\includegraphics[width=0.7\columnwidth]{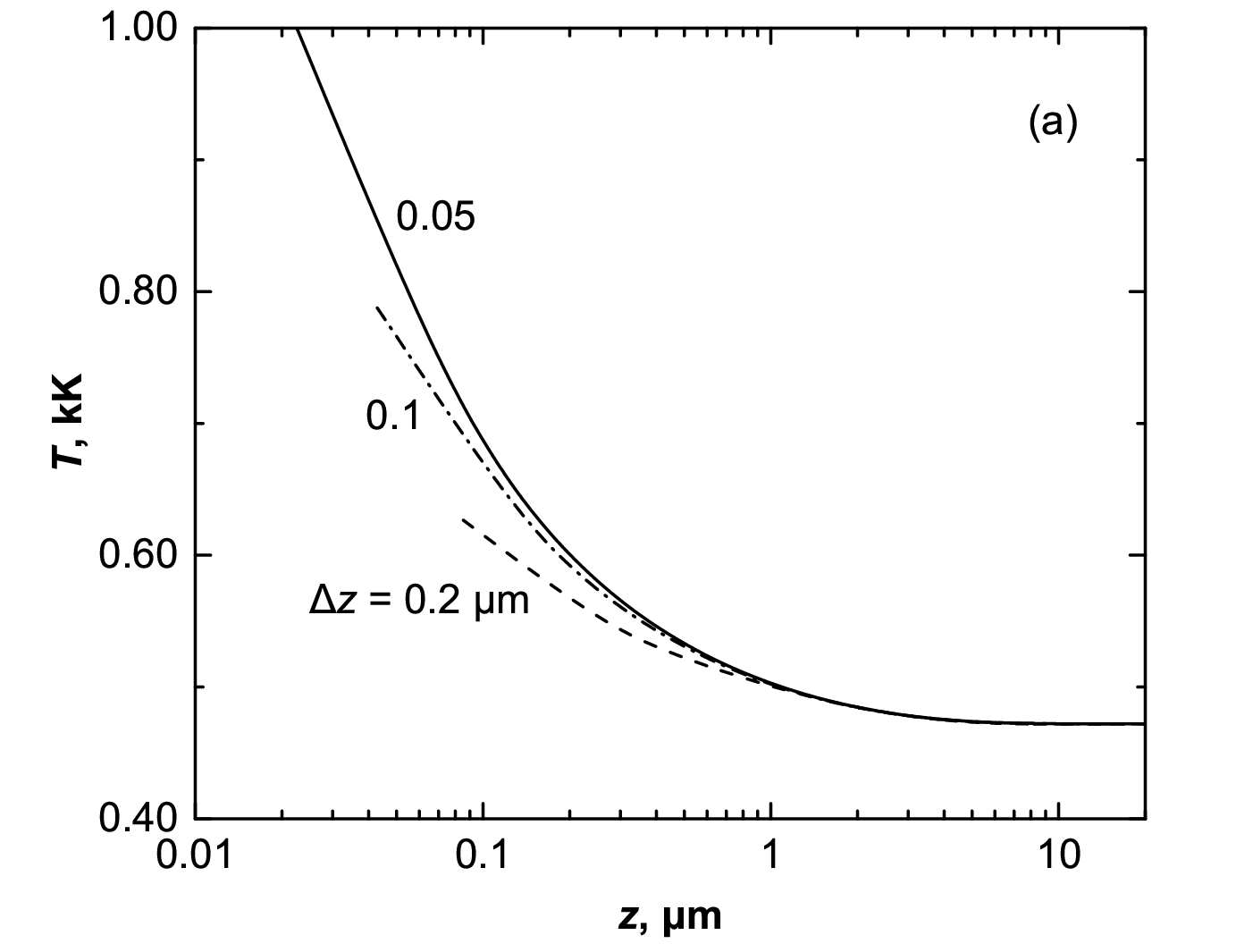}
\includegraphics[width=0.7\columnwidth]{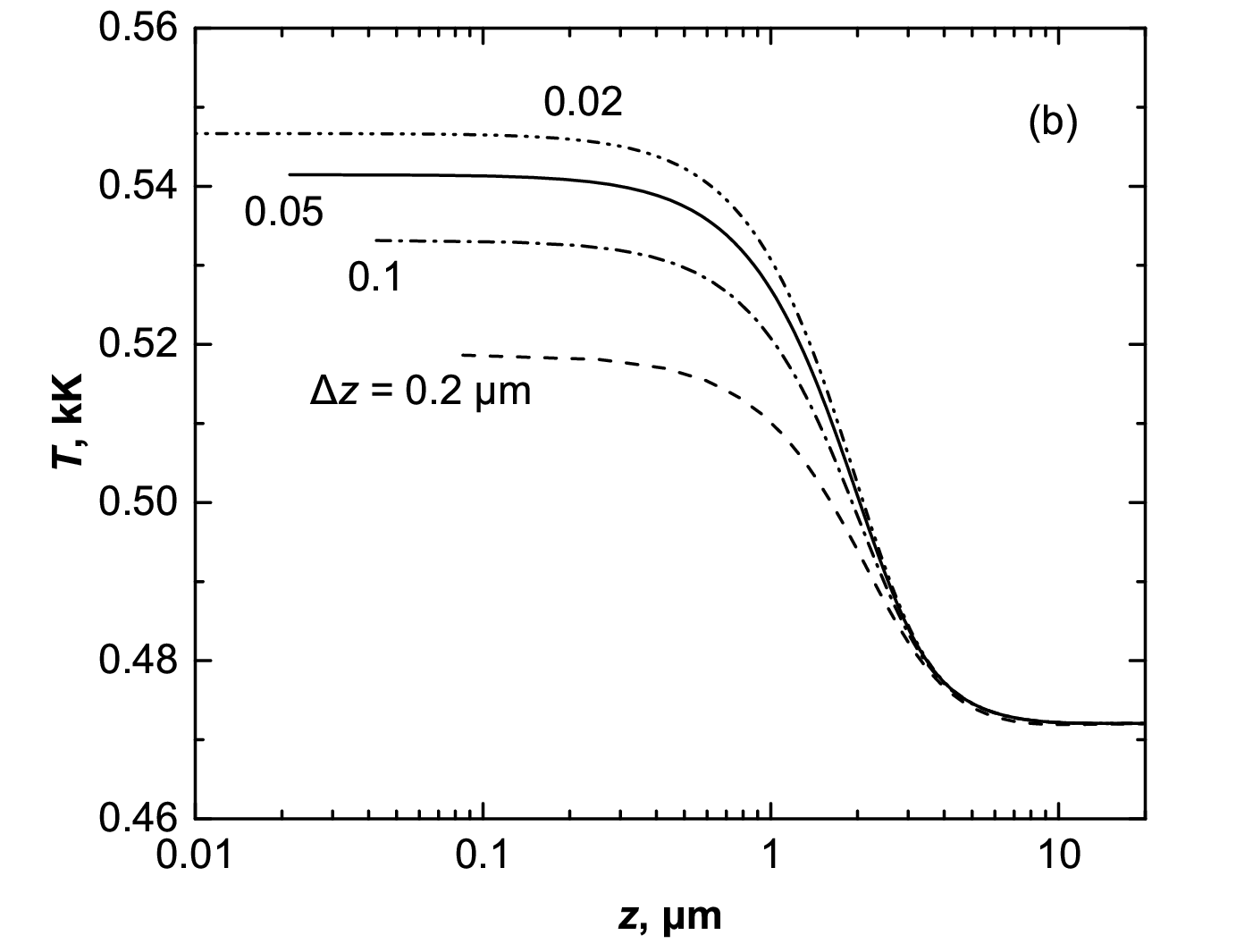}
\end{center}
\caption{\label{fig1}
High-entropy layer near the impact surface at symmetric collision of aluminum plates with velocity jump behind the shock wave $u=1$~km/s at $t = 10$~ns, $\eta^*=1$~Pa\,s from calculations in hydrodynamic approximation with different mesh size $\Delta z$ (a) without and (b) with taking into account the heat conductivity.
}
\end{figure}

The spatial distributions of the substance temperature, calculated in the model formulation (excluding thermal conductivity, $\kappa = 0$) are shown in Fig.~\ref{fig1}(a) for Al; velocity jump is $u=1$~km/s. The initial temperature value is $T_0=300$~K; the temperature behind the steady shock wave is $T_\mathrm{SW}\approx 470$~K; the substance temperature near the collision surface is higher, $T>T_\mathrm{SW}$: the high-entropy layer takes place. We observe this as the layer of higher temperature, but call that by the high-entropy layer because the entropy is commonly considered as the measure of non-equilibrium, which is the cause of the layer formation. The substantial temperature increase in comparison with $T_\mathrm{SW}$ takes place on distances about $h_\mathrm{SW}$ from the collision surface ($h_\mathrm{SW}\approx 0.6$~$\mu$m in the considered case). Numerical solution is at the limit of stability for the mesh size $\Delta z=0.2$~$\mu$m according to the condition of Eq.~(\ref{eq42}).

The total thickness of the high-entropy layer is independent upon the mesh size; that is determined only by the viscosity coefficient. The high-entropy layer profile (particularly the temperature distribution along spatial coordinate) at the distances higher than $h_\mathrm{SW}$ from the collision surface is also independent upon the mesh size. These are attractive properties of the obtained numerical solution distinguishing that one from the heating errors (or entropy traces) produced by artificial viscosity, as mentioned in Section~\ref{sec1}. It is evident in Fig.~\ref{fig1}(a) that a strong dependence of the temperature near the collision surface on the mesh size takes place. The finite-difference approximation of the strain rate is limited by the value of the order of $|\dot{\varepsilon}|<u/\Delta z$; therefore, the strain rate, the entropy and temperature near the collision surface are increasing with the decrease of $\Delta z$, nearly inversely to $\Delta z$ in accordance with Eq.~(\ref{eq2}). In the circumstances, an unlimited decrease of the mesh size leads to unlimited rise of temperature.

The calculation results with taking into account the heat conductivity are shown in Fig.~\ref{fig1}(b). The temperature near the collision surface is substantially lower than in the above-mentioned calculations illustrated by Fig.~\ref{fig1}(a), and that is not tend to infinity at $\Delta z \to 0$, quite the contrary, that converges to a finite limit, $T\approx 550$~K in the case. The heat conductivity also leads to the smearing of the high entropy layer with the lapse of time.

A stable profile of the high-entropy layer independent upon the mesh size can be obtained only with use of very fine computational grid ($\Delta z\approx 0.01$~$\mu$m in the considered case). It should be noted that perturbations of flow induced by non-parallelism or roughness of the impacting surfaces can influence on the high-entropy layer formation in the real conditions. Therefore, using of unreasonably fine grids can be senseless for most of the problems, for example, using of the mesh size $\Delta z$ less than the surface roughness.

According to data~\cite{Hildebrand-Lamoreaux-1976, Jakse-Pasturel-2013}, the shear viscosity of Al melt is about $\eta \sim 10^{-2}$~Pa\,s under normal pressure, while according to~\cite{Ceotto-Miani-HT-2013} it is even less, $\eta \sim 10^{-3}$~Pa\,s. Our simulations show that the stable solutions can be obtained at such low value of the viscosity coefficient $\eta^{*}\sim 10^{-2}$~Pa\,s, but with very small mesh size $\Delta z \approx 0.01$~$\mu$m. On the other hand, viscosity depends on pressure and temperature and, according to some estimations, can growth on orders of magnitude with the presuure increase on tens of gigapascals \cite{Mineev-Funtikov-PhysUsp-2004-eng}. Moreover, there is additional bulk viscosity $\zeta$, which contribute to the total dissipation in Eqs.~(\ref{eq2}) and (\ref{eq8}), but there is no information about its value. Obviously, the pressure-dependent shear and bulk viscosities of metals should be subjects for further investigations. In the present study, we use the viscosity coefficient as a parameter of calculations; this approach is validated in the next section.

\section{High-entropy layers in solid metals}
\label{sec6}
Like in the previous section ($\eta^* > 0$, $\kappa > 0$), here we solve the problem of symmetric collision of metal plates but with taking into account elastic-plastic properties of the material. The shock wave moves to the right from the collision surface; the substance behind the shock front is at rest, while the substance before the shock front moves to the left with velocity $-u$; velocity jump on the shock front equals $u$.

\begin{figure*}[t]
\begin{center}
\includegraphics[width=0.7\columnwidth]{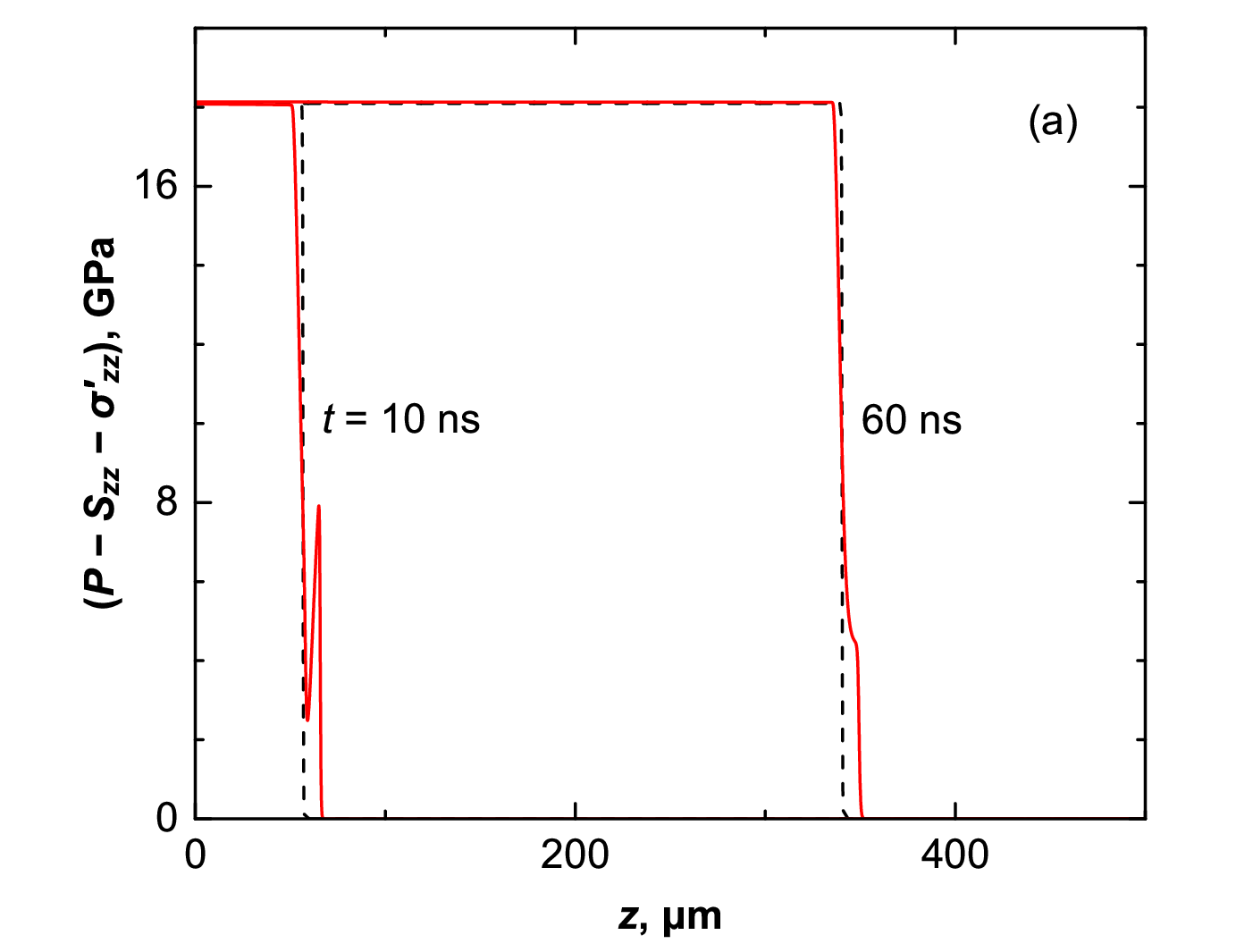}
\includegraphics[width=0.7\columnwidth]{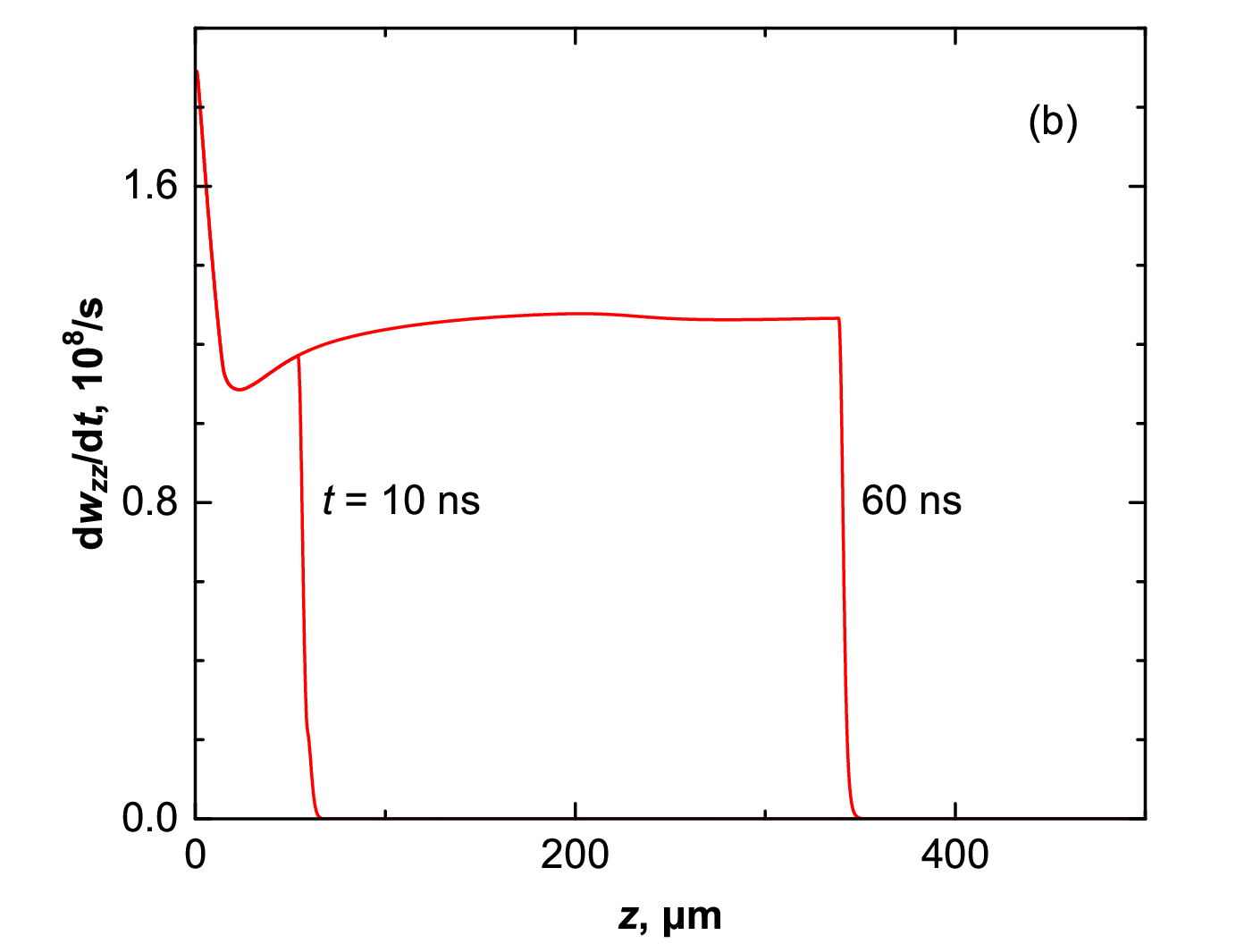}
\end{center}
\caption{\label{fig2}
Spatial distributions of (a) the longitudinal stress and (b) the maximal plastic strain rate at the symmetric collision of aluminum plates at $u=1$~km/s from elastic-plastic (solid line) and hydrodynamic (dashed line) simulations. The maximal plastic strain rate (b) is taken as a maximum in each point of substance over all time starting from the beginning of impact.
}
\end{figure*}

\begin{figure*}[t]
\begin{center}
\includegraphics[width=0.7\columnwidth]{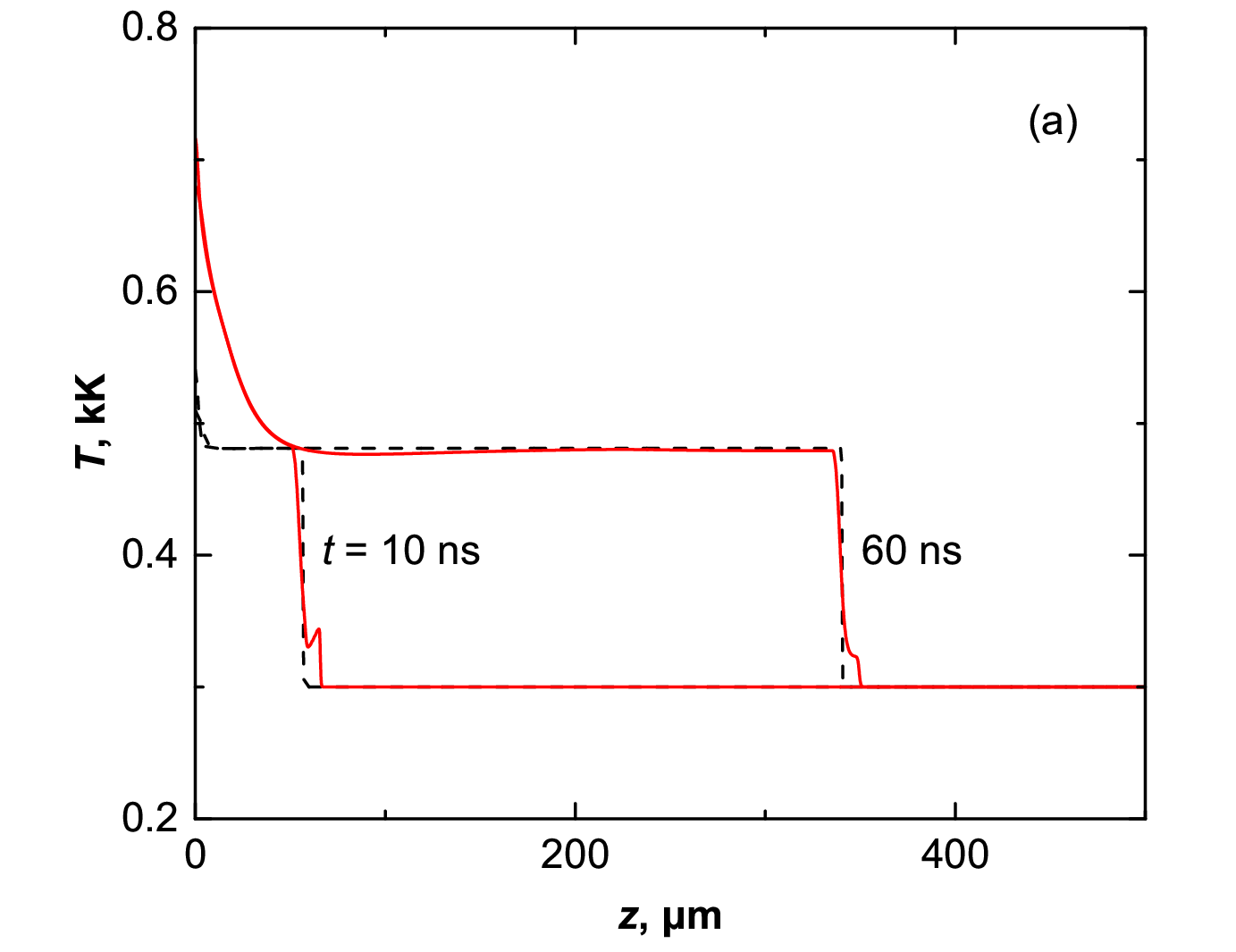}
\includegraphics[width=0.7\columnwidth]{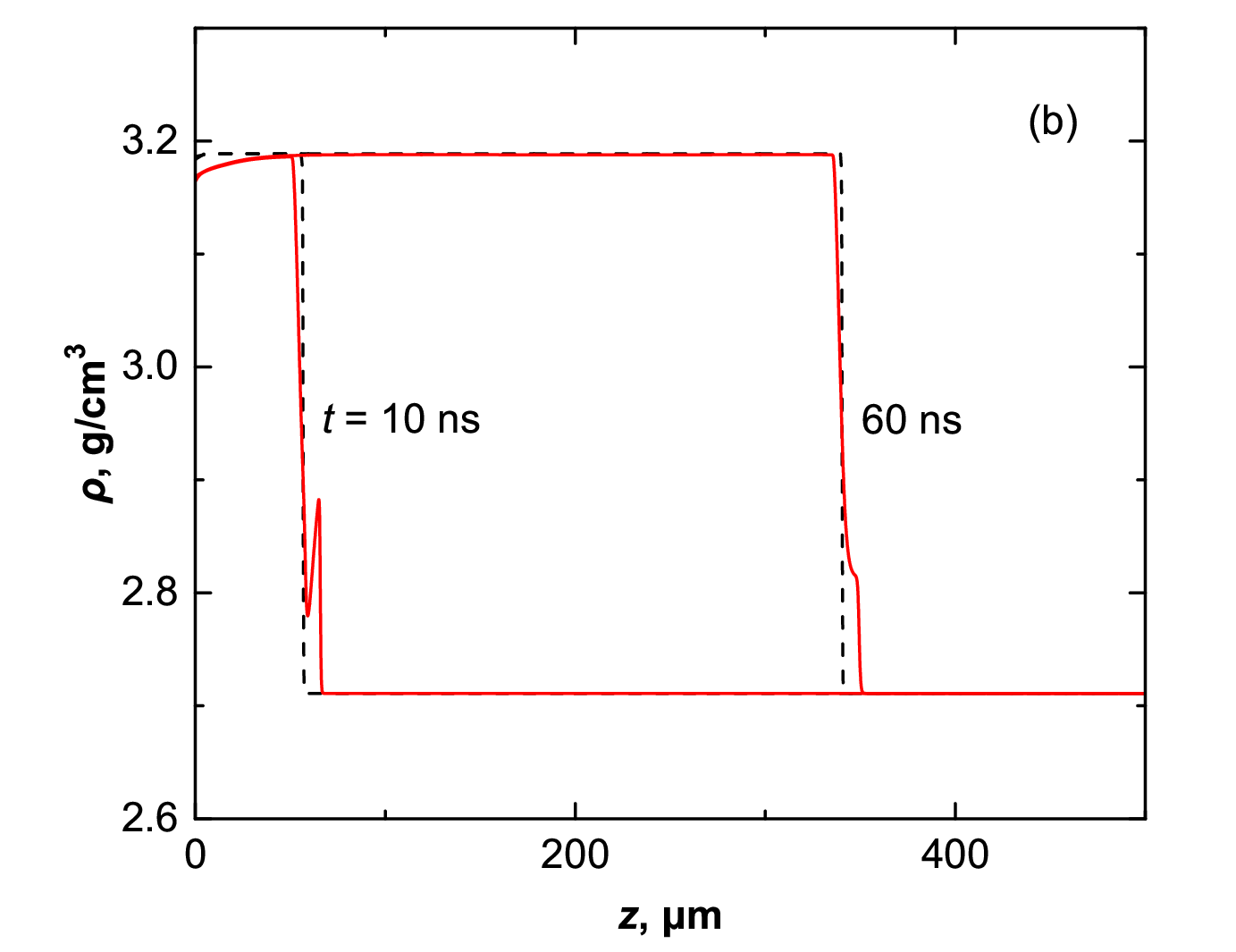}
\end{center}
\caption{\label{fig3}
Spatial distributions of (a) the temperature and (b) the density at the symmetric collision of aluminum plates at $u=1$~km/s from elastic-plastic (solid line) and hydrodynamic (dash line) simulations.
}
\end{figure*}

Comparison of results for symmetric collision of aluminum plates calculated in hydrodynamic and elastic-plastic approximations on numerical grid with the mesh size of $\Delta z\approx 0.1$~$\mu$m is made in Figs.~\ref{fig2} and~\ref{fig3}. The viscosity coefficient is $\eta^{*}=1$~Pa\,s, which value is enough to stabilize the numerical solution but weakly influence on the resulting temperature distribution. Accounting for the elastic-plastic properties results in the separation of a shock wave on the elastic precursor and the main plastic wave.

Far from the collision surface, the stress component, density and temperature behind the shock front tend to the certain constant values, which are nearly the same for both elastic-plastic and hydrodynamic calculations. 
In the case of elastic-plastic calculations, the temperature is slightly lower because a part of energy is stored in defects of crystalline structure (dislocations) and in the shear stresses. The sequence of energy transformations in the shock-compressed material is the following: compression pumps the elastic energy of the volume change, $U$, by means of work of pressure and the elastic energy of the form change, $U_\mathrm{s}$, by means of work of stress deviators; then plasticity transforms a part of energy $U_\mathrm{s}$ into $U$, which is the plastic dissipation. It should be emphasized, that we do not consider plasticity separately from elasticity, but operate with either hydrodynamic approximation, or elastic-plastic medium. If one compares purely elastic and elastic-plastic media, the temperature should be higher in the latter case, because a part of the energy stored in the shear stress field is released into the heat by means of plastic relaxation.

That contradicts to the data from~\cite{Swift-Seifter-Holtkamp-Clark-PRB-2007}, where an additional heating due to the plastic dissipation had been reported on the basis of calculations. This additional heating had been supposed by~\cite{Swift-Seifter-Holtkamp-Clark-PRB-2007} as a reason of the higher temperature detected by pyrometric measurements \cite{Swift-Seifter-Holtkamp-Clark-PRB-2007, Seifter-Swift-PRB-2008} in comparison with results of hydrodynamic calculations. We guess that the discussed discrepancy is explained by double accounting of the work of deviatoric stress~\cite{Swift-Seifter-Holtkamp-Clark-PRB-2007}: the first time in the equation for internal energy and the second time as additional heating due to the plastic work; see our comments to Eq.~(\ref{eq7}).

\begin{figure}[t]
\begin{center}
\includegraphics[width=0.7\columnwidth]{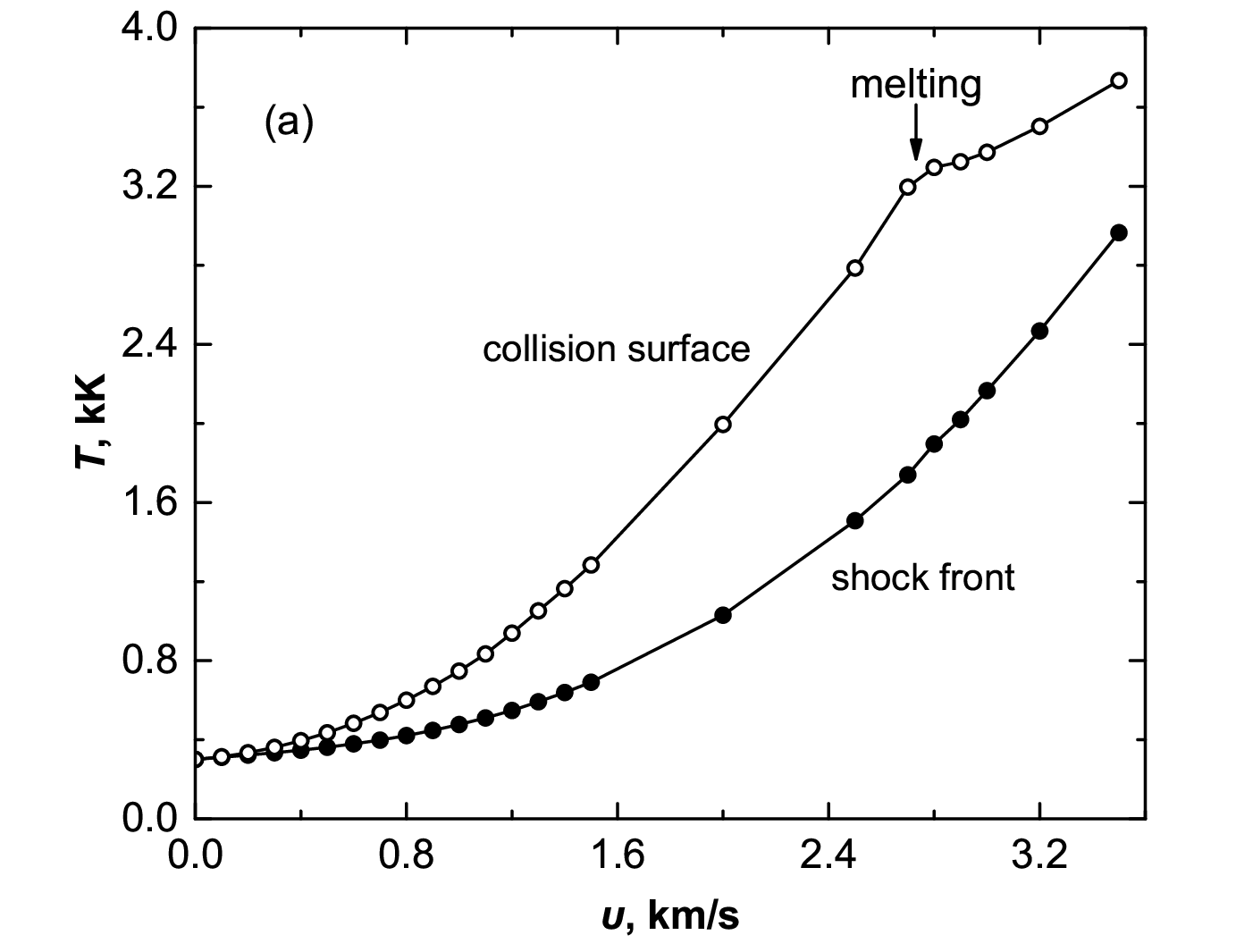}
\includegraphics[width=0.7\columnwidth]{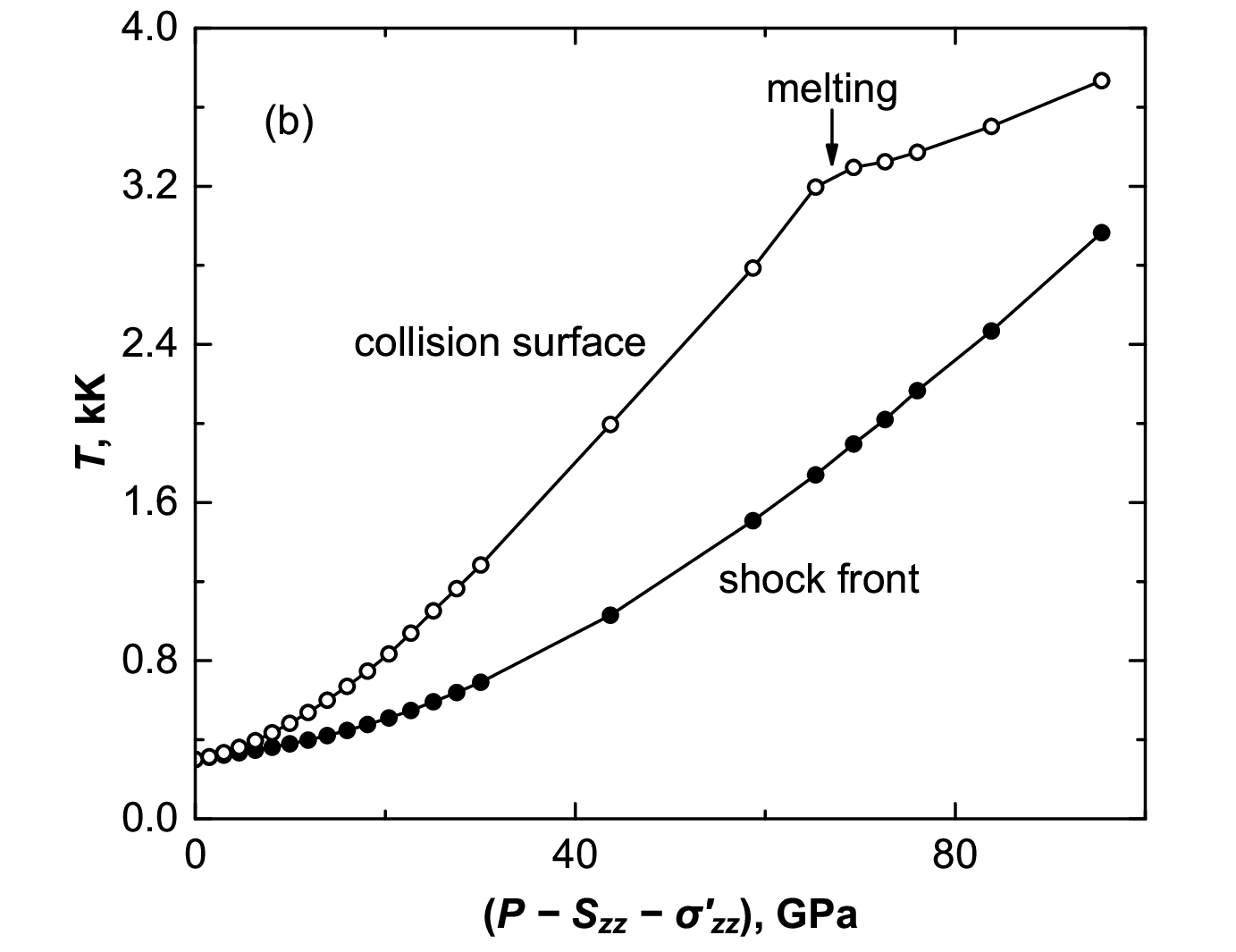}
\end{center}
\caption{\label{fig4}
Temperature on the collision surface in $t=10$~ns after impact and behind the shock wave front versus (a) the particle velocity jump in the shock wave and (b) the longitudinal stress behind the shock wave. The arrows indicate the beginning of melting near the collision surface just after the instant of impact $t=0$.
}
\end{figure}

Temperature distribution is inhomogeneous near the collision surface [see Fig.~\ref{fig3}(a)] with the maximum on the surface. The high-entropy layers reveal themselves as increasing of temperature [illustrated by Fig.~\ref{fig3}(a)] and decreasing of substance density [see Fig.~\ref{fig3}(b)] near the collision surface $z=0$ at the constant stress [see Fig.~\ref{fig2}(a)]. The high-entropy layer caused by plasticity is wider and more intensive than the corresponding layer in the case of viscous fluid; it means that the accounted physical viscosity weakly influence on the calculation results for the elastic-plastic medium. The high-entropy layer arises in the elastic-plastic medium due to the higher plastic strain rate near the collision surface in comparison with its value on the steady shock front [see Fig.~\ref{fig2}(b)].

Full width at half maximum of the high-entropy layer is about 16~$\mu$m in the case (a half of this width of about 8~$\mu$m is shown in Fig.~\ref{fig3} at $z>0$), that is close to the shock front thickness ($h_\mathrm{SW}\approx 11$~$\mu$m). At the viscous fluid approximation, one should use the viscosity coefficient of $\eta^*\approx 20$~Pa\,s to provide a comparable thickness of the shock wave front, see Eq.~(\ref{eq41}). The shock front thickness in solids should be determined by both viscosity and plasticity in general, but determination of physically based values of viscosity coefficient $\eta^*$ require an additional study as opposed to the plasticity effects, which can be calculated on the basis of self-consistent model~\cite{Krasnikov-Mayer-Yalovets-IJP-2011, Mayer-Khishchenko-Levashov-Mayer-JAP-2013}. 

In our present calculations, viscosity coefficient is free parameter and we tried to make its influence negligible, but it cannot be completely excluded because provides stability of a solution. Accounting for the plasticity solely can not provide the stability because that effects not similarly to viscosity. The viscosity makes an additional increase of compressive stresses in the substance elements those are compressing faster than other elements and prevents instability, while plasticity only relaxes the stresses. Therefore, determination of the viscosity coefficient in solids is an important topic for further investigations.

Calculated temperatures on the collision surface $z=0$ (in 10~ns after the impact) and behind the shock wave front versus the particle velocity and the longitudinal stress jumps on the steady shock wave are shown in Fig.~\ref{fig4}. The temperatures difference grows with the shock intensity increase at low stresses and can exceed 1~kK. As the pressure is constant behind the shock front, melting in the high-entropy layer begins at lower values of $u$ than after the steady shock wave in the bulk. Aluminum melts near the collision surface at $u>2.7$~km/s but remains solid in the bulk even at $u=3.5$~km/s. Subsequent increase of the loading intensity leads to decrease of the temperature difference between the high-entropy layer and the bulk (see Fig.~\ref{fig4}); it is because the substance melting in the high-entropy layer prevents further plastic dissipation and corresponding heating. The difference decrease leads to breaks on curves for the collision surface temperature in Fig.~\ref{fig4}; these breaks indicate the beginning of melting in the high-entropy layer near the collision surface (just after the impact).

\begin{figure}[t]
\begin{center}
\includegraphics[width=0.7\columnwidth]{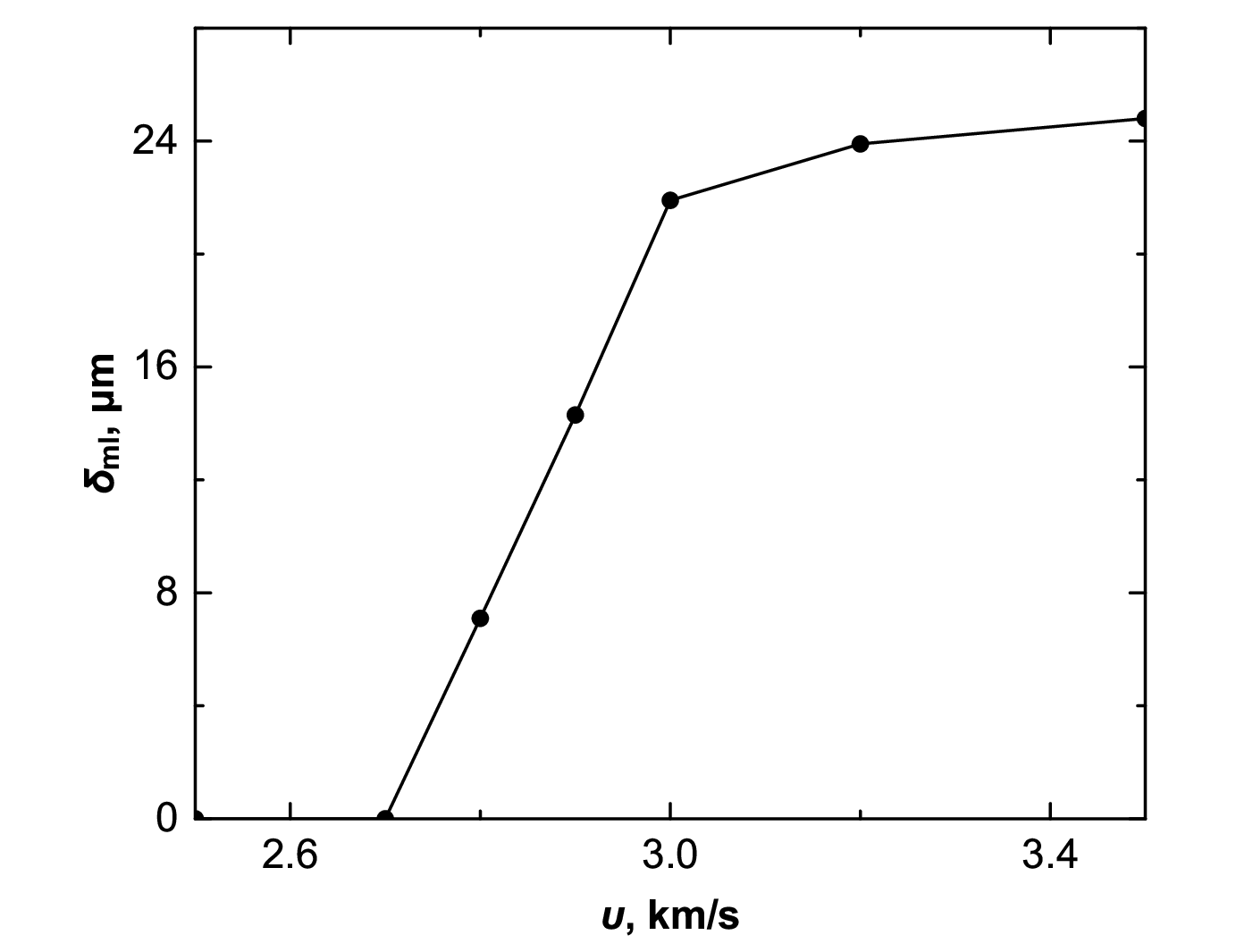}
\end{center}
\caption{\label{fig5}
Calculated maximal thickness of molten layer near the impact surface (in the high-entropy layer) at collision of aluminum plates as a function of the particle velocity jump in the shock wave.
}
\end{figure}

Calculated thickness $\delta_\mathrm{ml}$ of the molten layer near the collision surface that is the thickness of molten part of the high-entropy layer is shown in Fig.~\ref{fig5} as a function of the particle velocity jump $u$. After the beginning of melting, this thickness linearly grows with $u$ at first, then it becomes equal to the high-entropy layer thickness (about 25~$\mu$m) and ceases to grow. Such behavior corresponds to the temperature distribution in the high-entropy layer [see Fig.~\ref{fig3}(a)].

\section{Confirmation of high-entropy layers from MD simulations}
\label{sec7}
For validation of our prediction of the high-entropy layer near the collision surface, we perform MD simulations. Symmetric collision of two aluminum plates 0.5~$\mu$m thick each with free rear surfaces is considered. Simulation of two plates instead of one with mirror reflection increases the MD system size, but let us to ensure avoiding of an influence of artificial boundary condition at the collision surface. The system size in both transverse directions is only 0.02~$\mu$m with periodic boundary conditions in these directions; thus, we simulate a small piece of the colliding plates. The MD system contains about 21 million atoms. Each MD run takes from 35 to 50~h depending on the preparation stage with using 360 threads of supercomputer ``Tornado SUSU'' of the South-Ural State University (National Research University). Well-established and widely used MD simulator LAMMPS \cite{Plimpton-1995} is implemented with the angle-dependent (ADP) inter-atomic potential \cite{Apostol-Mishin-2011}, which is an extension of the embedded atom model (EAM) \cite{Daw-Baskes-1984} with accounting of the angular dependence of the interaction energy. Initial preparation stage includes different procedures for the considered samples, but in all cases it is finished by relaxation in the Nose--Hoover barostat and thermostat at zero stresses and temperature of 0.3~kK during at least 10~ps. After that, the collision velocity equal in absolute value and opposite in sign is imposed along the collision direction on the random thermal velocity of atoms of the two plates. The following evolution is calculated without thermo- and barostating at the constant energy of the MD system that simulates free collision of the plates. The temperature is calculated from the kinetic energy of atoms in the layers 0.007~$\mu$m thick containing about 150~thousands atoms each with previous deduction of the directed part of velocity, which is the average velocity of the group of atoms.

\begin{figure}[t]
\begin{center}
\includegraphics[width=0.7\columnwidth]{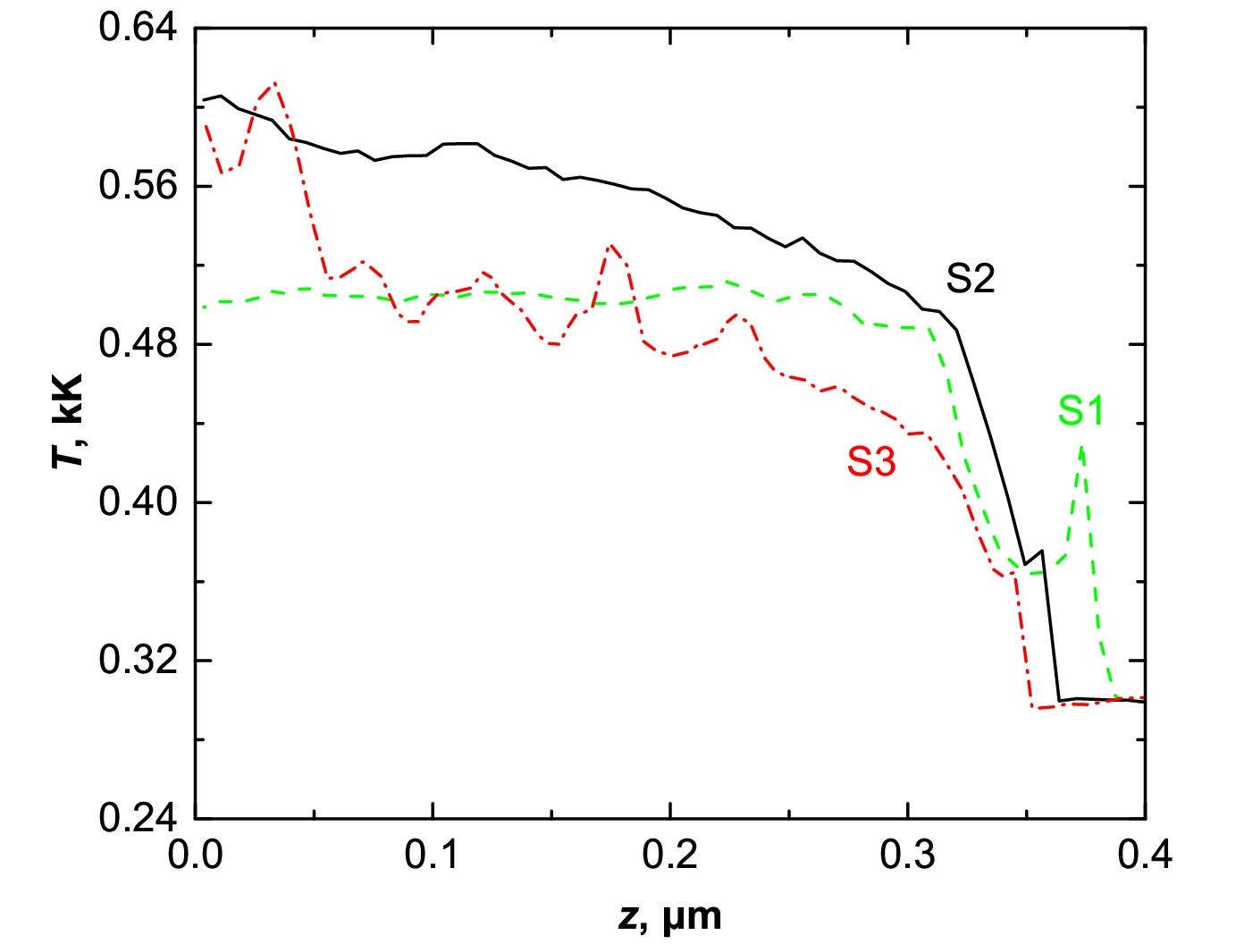}
\end{center}
\caption{\label{fig6}The temperature near the impact surface at symmetric collision of Al plates with different defectiveness (S1, S2 and S3) at velocity of 1~km/s in 60~ps after the impact moment obtained in MD simulations.
}
\end{figure}

According to our predictions from continuum model (see Fig.~\ref{fig3}), the thickness of the high-entropy layer in elastic-plastic case constitutes several tens of micrometers, which is not attainable for our MD simulations. The second difficulty is that, in MD, it is not possible to ``switch-off'' the shear stresses and consider hydrodynamic approximation. In order to resolve both problems, we consider three different initial states of the sample. The first one (S1) is a perfect single crystal with the lattice direction [100] coinciding with the collision direction. In this case, there are no carriers of plastic deformation in the origin material, and the plasticity incipience and development is delayed. As one can see from the comparison of Figs.~\ref{fig3}(a) and \ref{fig6}, the sample S1 is the most close to an ideal medium. In the case of collision with velocity of 1~km/s, the temperature distribution behind the shock wave front forms a plateau near 0.5~kK. This is quite close to the continuum results far from the collision surface [see Fig.~\ref{fig3}(a)]. Thus, we use S1 as a reference sample reflecting behavior far from the collision surface. It should be also mentioned that the longitudinal stress behind the shock wave front in MD simulations is about 18~GPa, which is quite close to the continuum results [see Fig.~\ref{fig2}(a)].

\begin{figure}[t]
\begin{center}
\includegraphics[width=0.7\columnwidth]{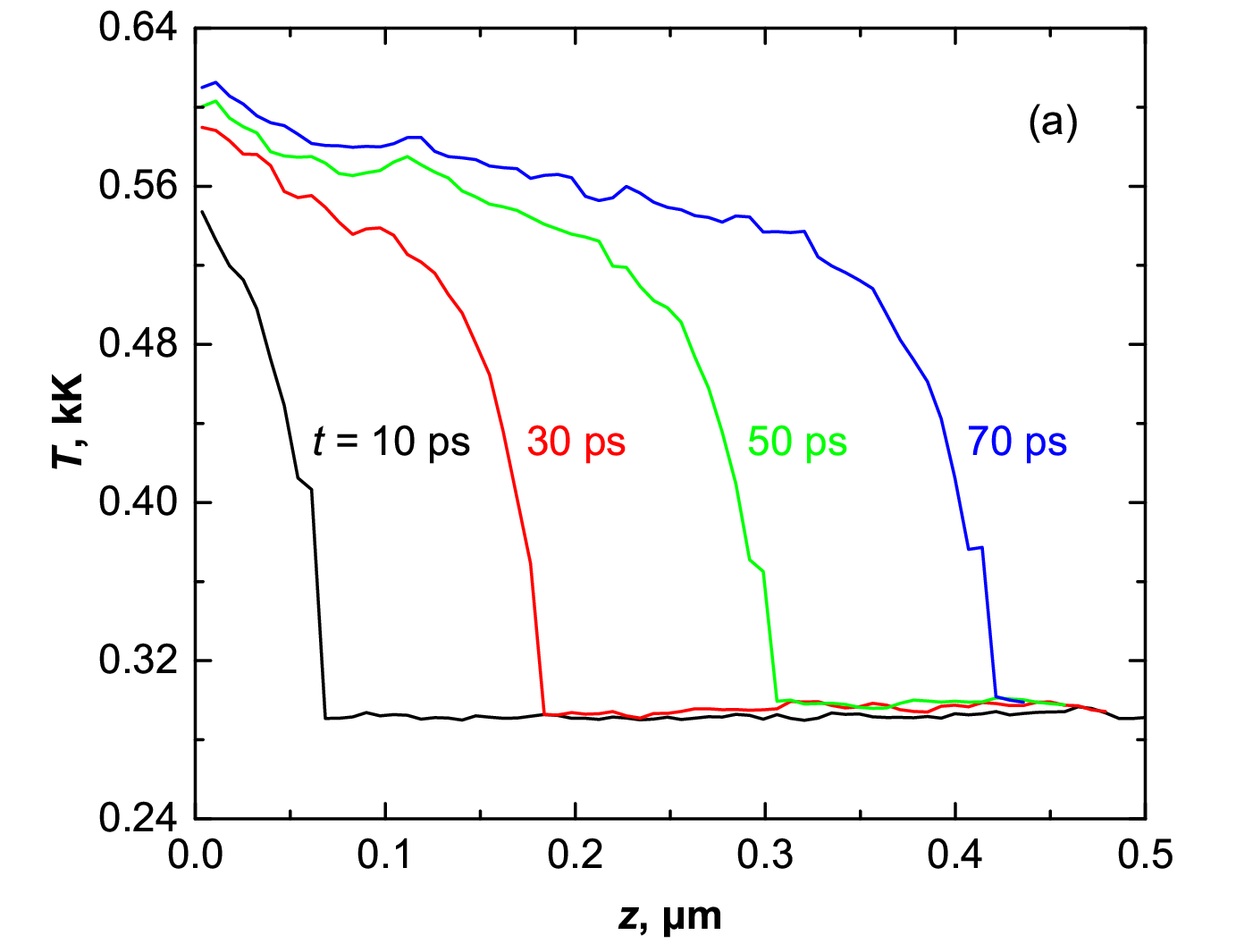}
\includegraphics[width=0.7\columnwidth]{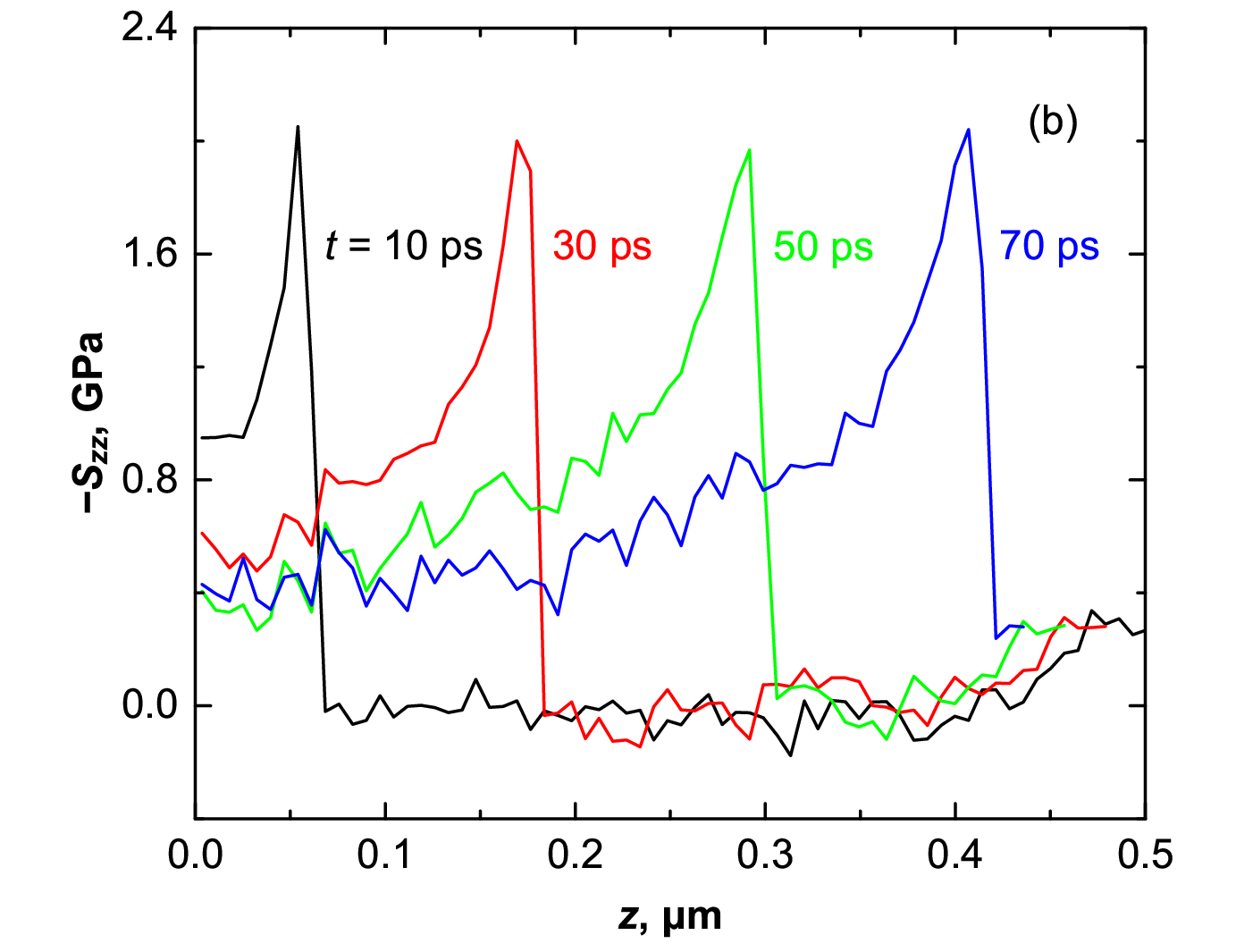}
\end{center}
\caption{\label{fig7}Time evolution of (a) the temperature field and (b) the stress deviator field for the case of samples with moderate defectiveness (S2) at symmetric collision of Al plates with velocity of 1~km/s from MD simulations. Zero time corresponds to the instant of impact.
}
\end{figure}

The second sample (S2) has a moderate initial defectiveness of lattice, which is formed by the following preparation procedure. Initial material with porosity of 0.1 is created (random deleting of 10\% of atoms) and uniaxially compressed in a direction perpendicular to the future collision direction up to the deformation of 0.15 during 15~ps. This deformation of porous metal initiates plastic collapse of pores and produces a solid sample with lattice defects. A collateral result of the preparation is excitation of stress waves due to free boundary conditions on the plate surfaces. For dumping of the emerging vibrations, a 7-step procedure of equilibration is implemented with zeroing atom velocities followed by an assignment of random thermal velocities and holding in the barostat and thermostat for 5~ps in each step. As one can see in Fig.~\ref{fig6}, the sample S2 with initially defective lattice demonstrate an almost 0.1~kK higher temperature near the collision surface in comparison with the ``ideal'' S1 sample and the prediction of continuum model far from the collision surface [see Fig.~\ref{fig3}(a)]. Also one can see that the temperature decreases with approaching of the shock wave front in the case of S2 sample. Both these facts evidence the formation of the high-entropy layer near the collision surface for the defective sample prone to elastic-plastic behavior. Time evolution of the temperature field is presented in Fig.~\ref{fig7}(a). The temperature increase continues in the shocked material due to gradual plastic release of the energy accumulated in the shear stresses characterized by the stress deviator [Fig.~\ref{fig7}(b)]. It confirms the used energy balance in continuum model, which takes into account initial energy accumulation in the shear stresses with the following plastic release leading to the increase of temperature [see Eq.~(\ref{eq7}) and the related discussion]. Analysis of the temperature evolution shows that there is no approaching to a stationary level of temperature behind the shock wave for the considered distances from the collision surface. This corresponds to the continuum model predictions of the thickness of high-entropy layer many times exceeding the size of the MD system.

The third sample (S3) with high initial defectiveness is prepared similar to S2, but with the initial porosity of 0.2 and uniaxial compression up to the deformation of 0.3 during 30~ps. As the result, S3 is characterized by almost completely destroyed crystal lattice and close in behavior of high-entropy layer to the case of highly viscous fluid. Namely, there is a narrow high-entropy layer with temperature approaching the reference level of 0.5~kK at the distance about 0.1~$\mu$m from the collision surface. A characteristic feature is also a high level of oscillations in the spatial distribution of temperature connected with non-uniform plastic deformation originated from non-uniform initial defect structure. We can conclude that the intensity of high-entropy layer depends on the initial defectiveness of the sample, which is characterized by the initial level of dislocation density in the continuum model and controlled by initial porosity in the case of MD simulations. The difference in the peak value of temperature between the MD (about 0.6~kK for S2) and continuum model (about 0.7~kK) can be attributed to the different initial levels of defect concentration. In the continuum model, we consider a typical value of defectiveness. Investigation of the dependence of the intensity of high-entropy layer on the defectiveness is a separate topic beyond the current research.

\section{High- and low-entropy layers at~ramp loading}
\label{sec8}
Ramp loading is one of the widespread experimental techniques; it often takes place at laser irradiation, for example. Here we consider the ramp loading within the frames of continuum model. The ramp loading is modeled as a linear increase of pressure on loaded surface during the rise time $\tau$, after then the pressure is constant:
\begin{equation}
P(t)=\begin{cases}
P_\mathrm{m} t/\tau, & t<\tau,\\
P_\mathrm{m}, & t\geqslant\tau,
\end{cases}
\label{eq43}
\end{equation}
where $P_\mathrm{m}$ is the maximal acting pressure. A metal plate is considered; the front surface is loaded, while the rear surface is free; the substance is at rest initially.

\begin{figure}[t]
\begin{center}
\includegraphics[width=0.7\columnwidth]{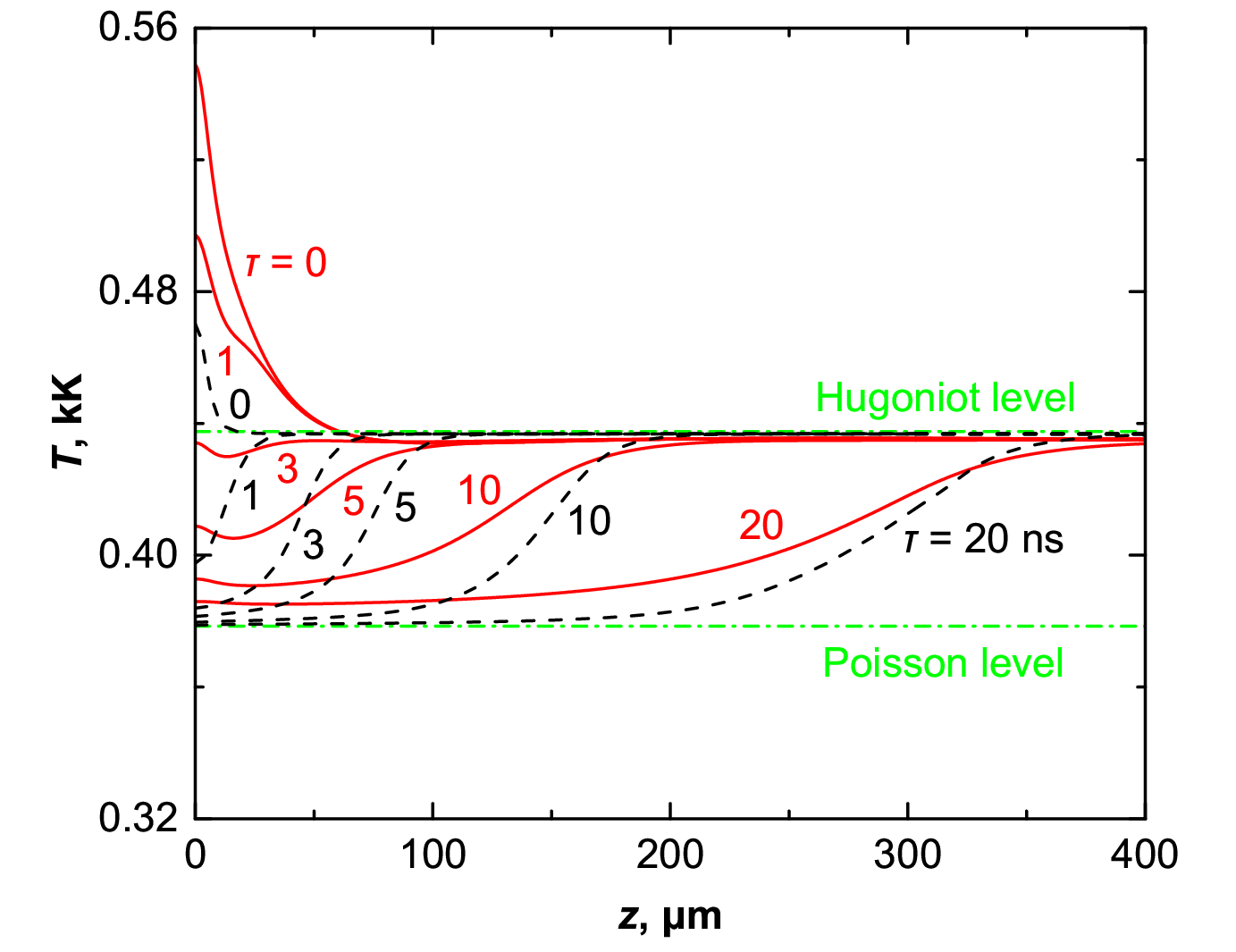}
\end{center}
\caption{\label{fig8}
Temperature profiles in high- and low-entropy layers near the loaded surface ($z=0$) at ramp loading of aluminum with the maximal pressure $P_\mathrm{m}=15$~GPa and various rise time $\tau = 0$, 1, 3, 5, 10 and 20~ns from elastic-plastic (solid lines) and hydrodynamic (dash line) simulations. 
}
\end{figure}

\begin{figure}[t]
\begin{center}
\includegraphics[width=0.7\columnwidth]{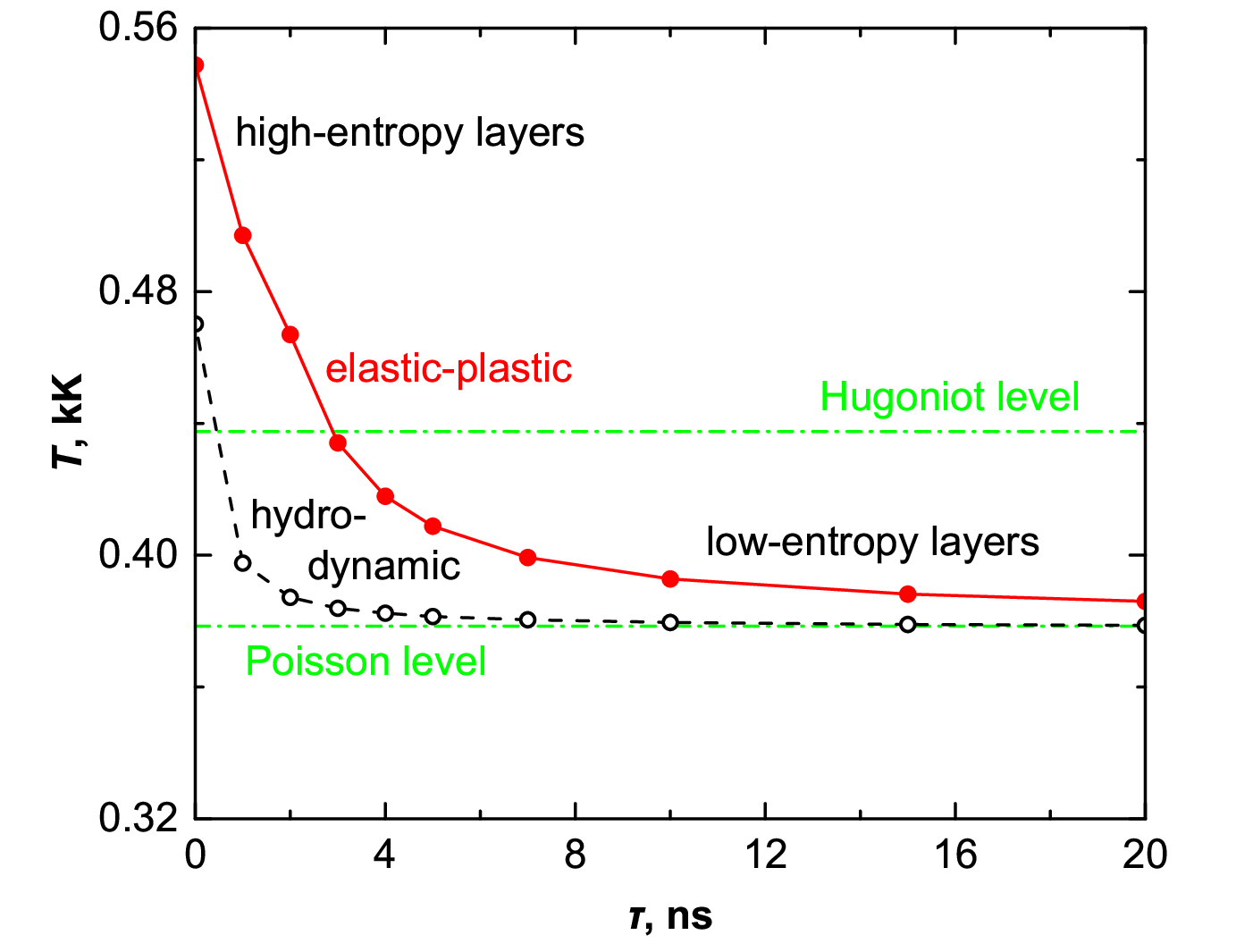}
\end{center}
\caption{\label{fig9}
The loaded surface temperature of aluminum at the loading pressure $P_\mathrm{m}=15$~GPa as a function of the rise time from elastic-plastic (solid line, solid circles) and hydrodynamic (dash line, open circles) simulations.
}
\end{figure}

\begin{figure}[t]
\begin{center}
\includegraphics[width=0.7\columnwidth]{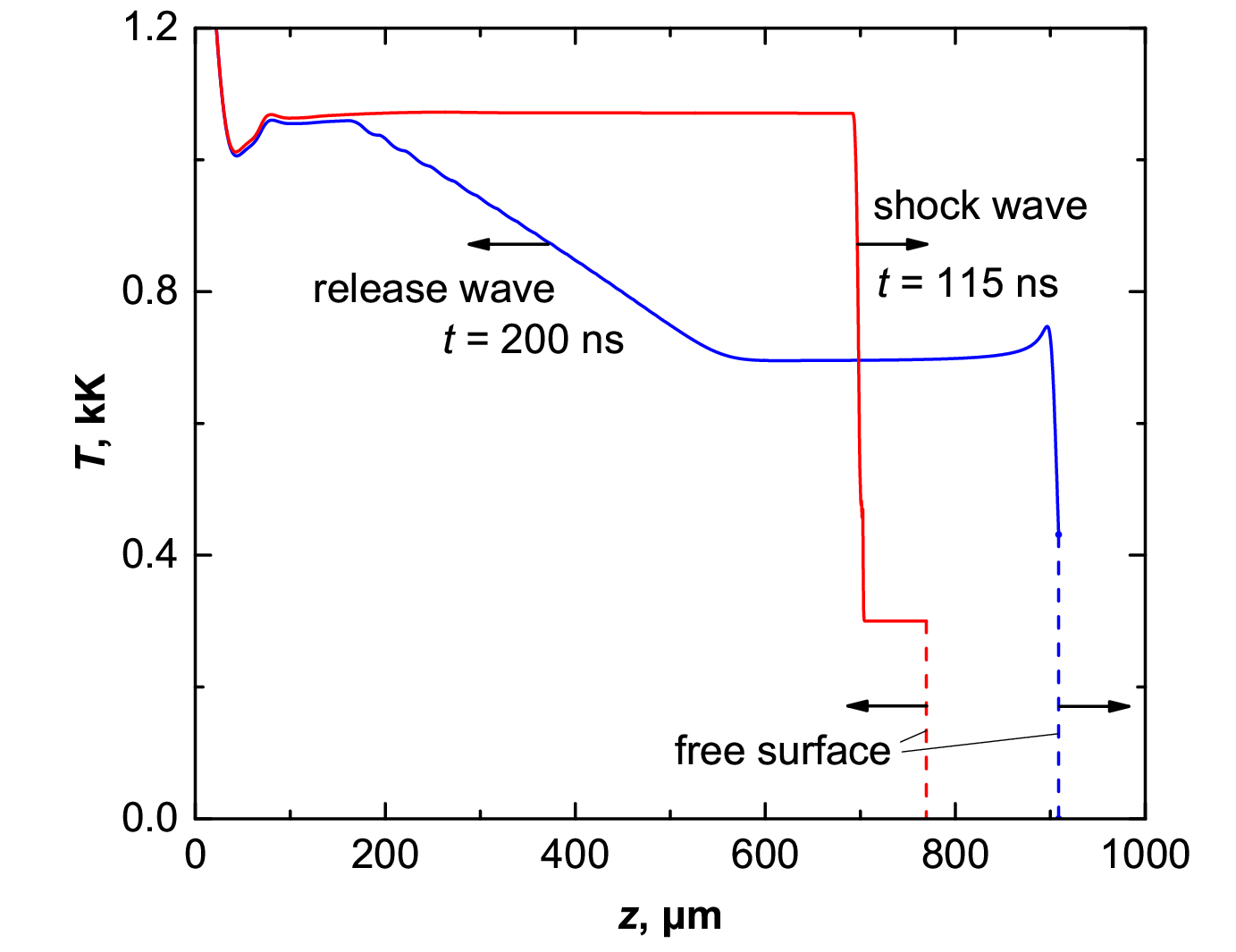}
\end{center}
\caption{\label{fig10}
Spatial distribution of the temperature in aluminum under the incident shock wave with $u=2$~km/s and behind the release wave reflected from the sample free surface.
}
\end{figure}

\begin{figure}[t]
\begin{center}
\includegraphics[width=0.7\columnwidth]{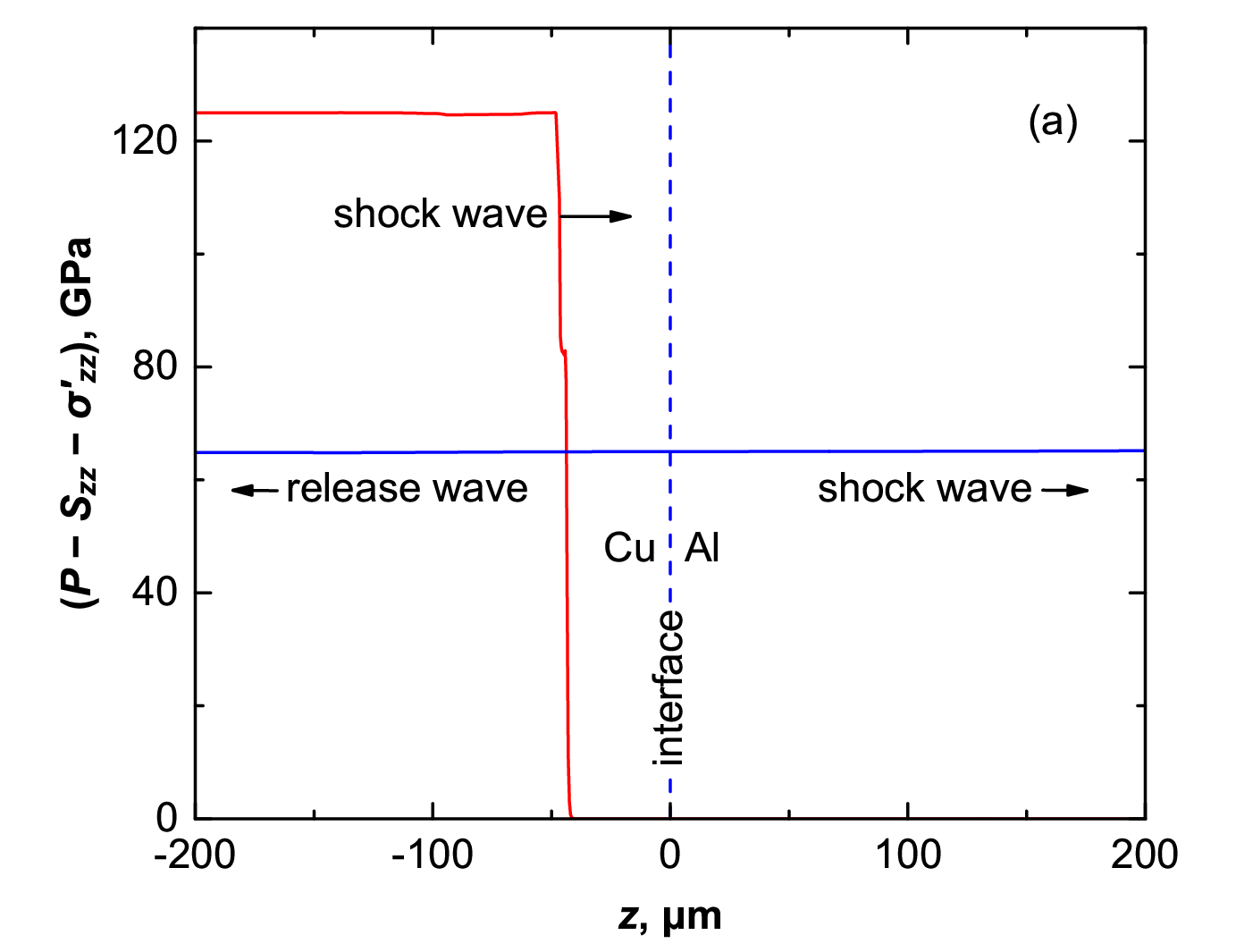}
\includegraphics[width=0.7\columnwidth]{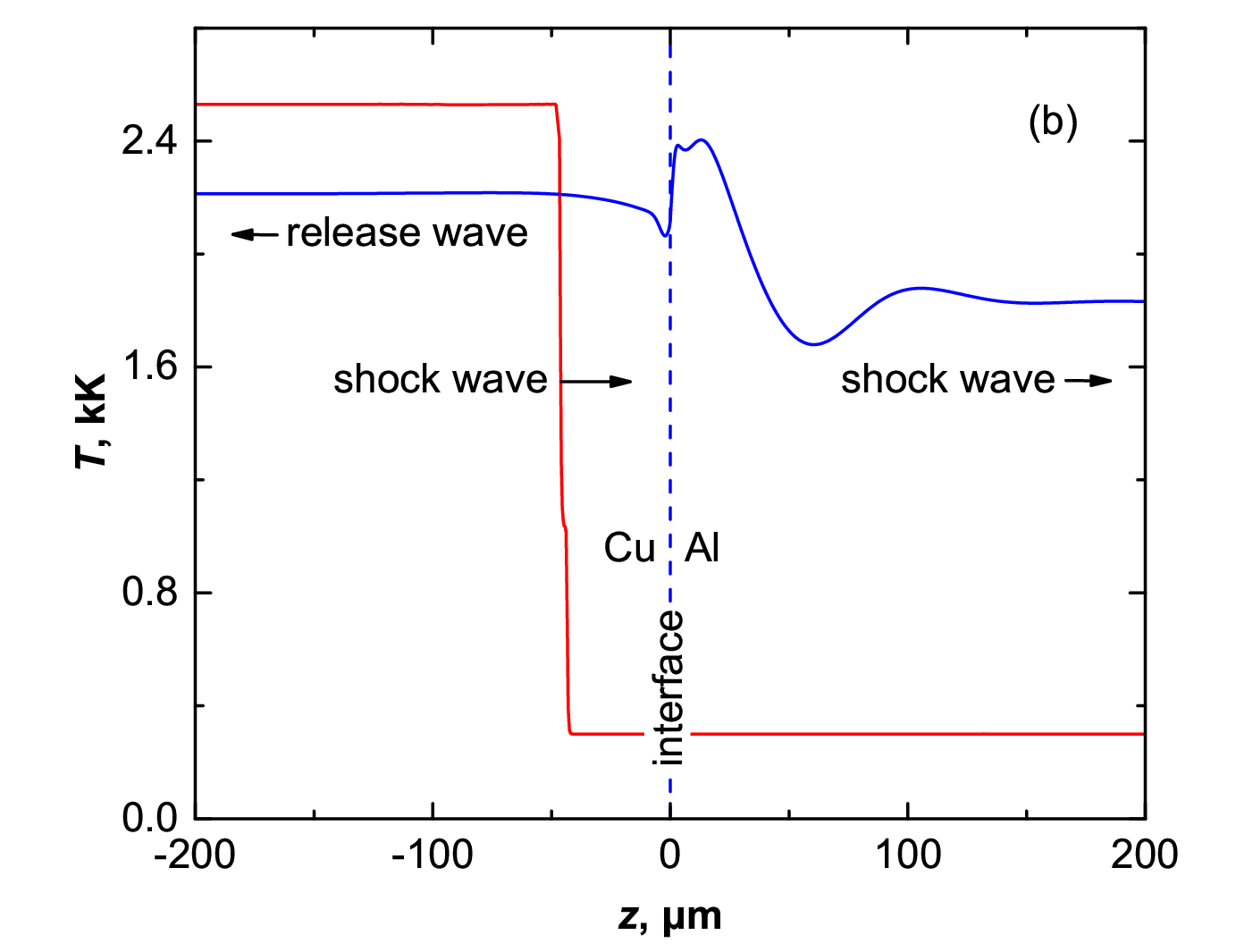}
\end{center}
\caption{\label{fig11}Stress (a) and temperature (b) distribution near the interface (dashed line) between copper and aluminum before (red solid line) and after (blue solid line) the incidence of the shock wave with $u=2$~km/s propogated from Cu to Al.
}
\end{figure}

\begin{figure}[t]
\begin{center}
\includegraphics[width=0.7\columnwidth]{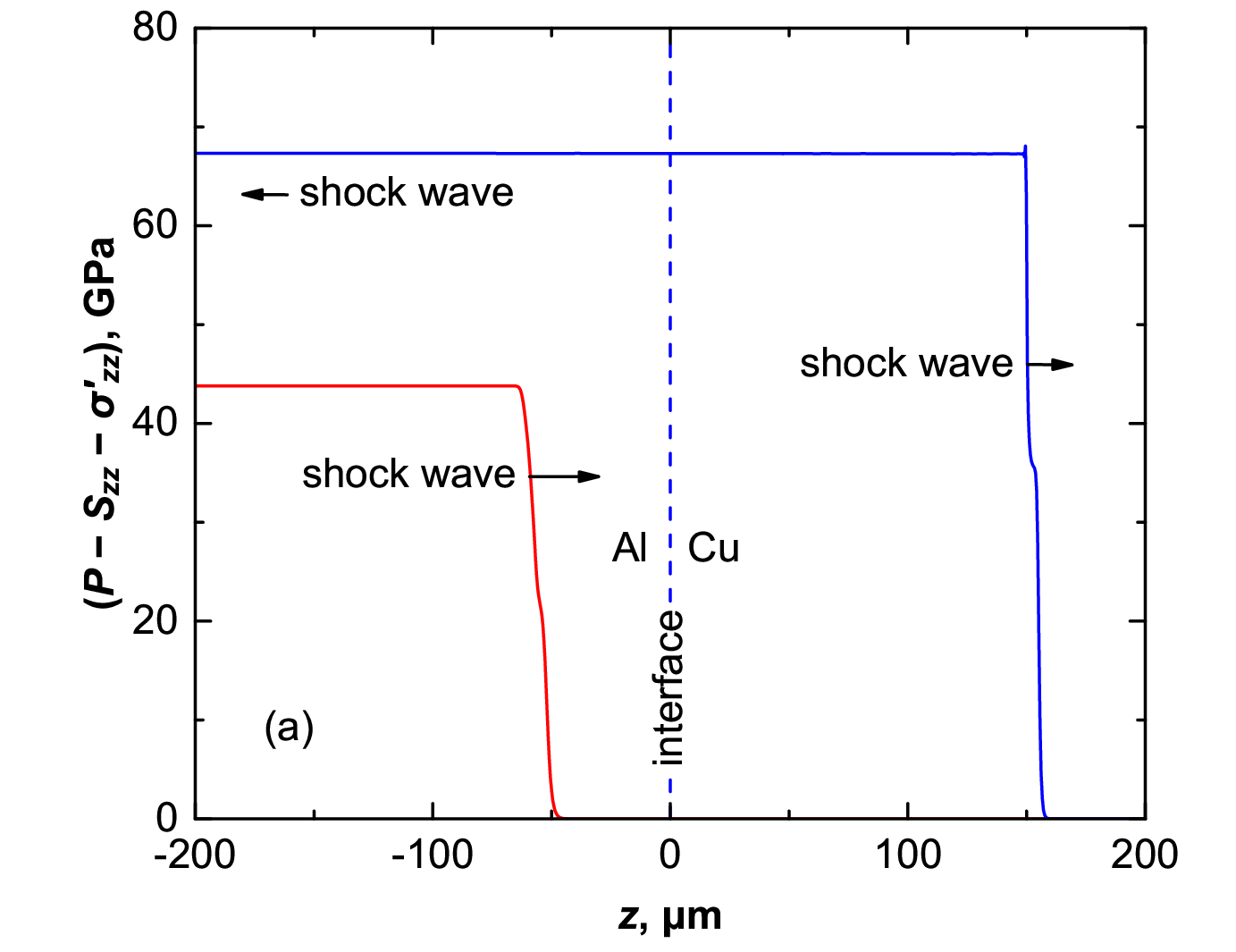}
\includegraphics[width=0.7\columnwidth]{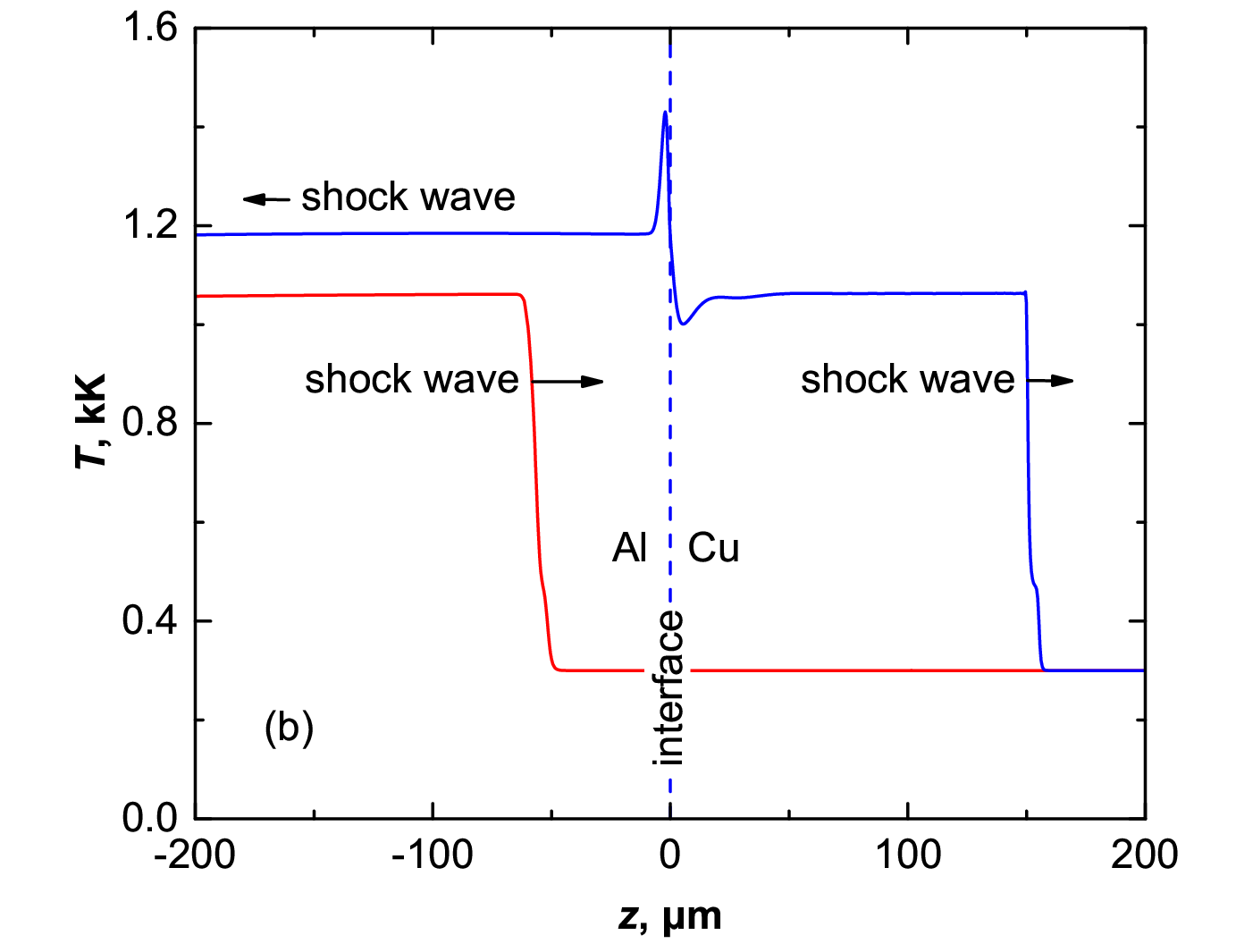}
\end{center}
\caption{\label{fig12}Stress (a) and temperature (b) distribution near the interface (dashed line) between aluminum and copper before (red solid line) and after (blue solid line) the incidence of the shock wave with $u=2$~km/s propogated from Al to Cu.
}
\end{figure}

Temperature fields near the loaded surface are presented in Fig.~\ref{fig8} for the case of aluminum loading with the maximal pressure $P_\mathrm{m}=15$~GPa. Various rise times are considered; zero rise time corresponds to the shock loading (like at the plate impact). At any loading rise time, an initially unsteady compression wave transforms into the steady shock wave with the lapse of time. After this transformation, the temperature behind the compression wave becomes constant (the steady shock wave level in Fig.~\ref{fig8}). Near the loaded surface the temperature is higher or lower than the steady shock wave level, so the high- or low-entropy layers are formed depending on the rise time. If the rise time is small, initial strain rate is higher than on the steady shock front, and the high-entropy layer is formed; limiting case $\tau=0$ corresponds to the results of Section~\ref{sec6}. On the contrary, large rise time leads to a lower initial value of the strain rate, and the low-entropy layer is formed; this situation is more common for the ramp loading. The rise time $\tau=3$~ns is approximately equal to the thickness (in time units) of the steady shock front, and almost flat temperature distribution is formed as a result (see Fig.~\ref{fig8}).

Results of hydrodynamic calculations are also presented in Fig.~\ref{fig8} for comparison. In the hydrodynamic calculations, the temperature tends to the Hugoniout adiabat level far from the loaded surface, as well as it tends to the Poisson adiabat level near the loaded surface with increase of the rise time $\tau$. The temperature behind the steady shock for elastic-plastic calculations is slightly lower than in the hydrodynamic approximation and than the Hugoniot adiabat level due to expenditure of energy on the formation of new dislocations, as it is mentioned in Section~\ref{sec6}. Near the loaded surface, the temperature is, vise versa, higher for the elastic-plastic calculations due to additional dissipation mechanism.

The rise time dependence of the loaded surface temperature is presented in Fig.~\ref{fig9}. It is monotonically decreasing with the maximal value corresponding to the zero rise time. This dependence crosses the Hugoniot adiabat level at $\tau \approx 3$~ns and tends to the Poisson adiabat level with increase of the rise time; thus, compression in the low-entropy layer becomes close to the isentropic one at $\tau \to \infty$ \cite{Hayes-Hall-Asay-Knudson-JAP-2004, Ding-2006, Ding-Asay-JAP-2007}. Extension of the rise time also increases the thickness of the low-entropy layer.

\section{Shock wave interaction with~interfaces}
\label{sec9}
Non-uniform distribution of temperature can be expected near various interfaces after the shock wave interaction with them. The reason is the evolution of the shock (or release) wave structure at crossing of the interface. At the evolution, the strain rate, the entropy production and the heating vary from point to point. 

The temperature field near the rear free surface of aluminum plate after reflection of the shock wave with velocity jump $u=2$~km/s is shown in Fig.~\ref{fig10}. Far from the rear surface, the temperature tends to a constant level corresponding to the action of steady shock and release waves. Near the surface the distribution of temperature is considerably inhomogeneous. It has a maximum on depth of about 12~$\mu$m from the rear surface with temperature increase about 50~K in comparison with deeper layers, while on the surface itself the temperature drops on about 300~K. This temperature drop occurs because the surface layer does not undergo the complete compression as the deeper layers do: due to the finite thickness of the shock front, unloading starts here before the compression is finished. Thus, the low-entropy layer is formed mainly, while the high-entropy layer (near the temperature maximum) is also exists. The last one is formed during the evolution of the unloading wave structure.

Interaction of the shock wave with interface between two different materials is illustrated in Figs.~\ref{fig11} and~\ref{fig12} by the example of copper and aluminum. The case of Fig.~\ref{fig11} corresponds to the shock wave crossing from harder material into the softer one, where the reflection of unloading wave happens. While in the case of Fig.~\ref{fig12}, the harder and softer materials are interchanged, and the reflection of shock wave takes place. Temperatures far from the interface correspond to the action of steady shock and release waves. Near the interface, the low-entropy layer is formed in the harder material (Cu), and the high-entropy layer is formed in the softer material (Al). Structure of these layers can be more complicated than at the considered above shock or ramp loading.

It is remarkable that the incident shock waves in Figs.~\ref{fig11}(a) and~\ref{fig12}(a) are characterized by strong elastic precursors, the shear stresses in which exceed the ideal shear strengths of Cu and Al at zero pressure, respectively. The substance states with such shear stresses can exist at high pressures, because both the shear modulus and the shear strength related to the modulus grow together with pressure. The shear stress is restricted by multiplication of dislocations, which is taken into account in the applied model, and by homogeneous nucleation of dislocations~\cite{Norman-Yanilkin-PSS-2012}. Molecular-dynamics simulations~\cite{Mayer-Krasnikov-Pogorelko-ICTAEM-2019} of simple shear of Cu single crystals show that the nucleation threshold grows from 10 to 30~GPa at the pressure growth from zero to 50~GPa. Thus, there is no condition for homogeneous nucleation of dislocations in the considered case, because the maximal achieved shear stress in Cu is 22~GPa at the pressure of 80~GPa [see Fig.~\ref{fig11}(a)]. In the case of Al, the situation is similar.

Existence of high- and low-entropy layers near an interface can substantially influence on the pyrometry measurement of temperature as that analyzes radiation from a substance layer near the interface (for example, near free surface or interface between the studied substance and a window), where the temperature differs from that of bulk \cite{Swift-Seifter-Holtkamp-Clark-PRB-2007}.
Moreover, the high- and low-entropy layers should be taken into account in such problems, where surface plays an important role, particularly, at loading of thin targets.

\section{Conclusions}
\label{sec10}
Theoretical analysis and computations show that the high- and low-entropy layers arise near the collision surface and other interfaces in problems of shock and ramp loading. The cause is the formation or transformation of the steady shock wave structure. At impact, the initial strain rate, entropy production and heating are higher near the collision surface than on the steady shock front; therefore, the high-entropy layer is formed. The distance passed by the compression wave after the impact before its conversion into the steady shock wave is comparable with the shock front thickness, which is controlled by viscosity (in fluids and solids) as well as plasticity (in solids). The formation of high-entropy layer at impact and the energy pumping from the shear stresses into the heat incorporated into the continuum model are supported by MD simulations. Ramp loading can form the high- or low-entropy layers near the loaded surface depending on the pressure rise time, but the low-entropy layer is the most expected. Reflection of the shock wave from a free surface produces a considerable low-entropy (in general) layer near the surface; the particular reason here is that the surface layer does not undergo the complete compression as the substance in the bulk does: unloading starts here before the compression is finished due to the finite thickness of the shock front. At the shock wave crossing the interface between two substances, the low-entropy layer is formed in the higher-impedance material, and the high-entropy layer is formed in the lower-impedance material.

The high- and low-entropy layers should be accurately taken into account at calculations of the thin targets loading and other situations where surface is important. Our calculations show that the surface temperature can be twice as high as the bulk temperature. For example, melting can occur in the high-entropy layer, while the substance remains solid behind the steady shock wave deeper in the bulk. One more example occurs at the pyrometry measurements, where the temperature perturbations near the examined surface can affect the result. High-entropy layer near the sample surface can be responsible for the increased temperature detected by pyrometric measurements~\cite{Swift-Seifter-Holtkamp-Clark-PRB-2007, Seifter-Swift-PRB-2008} compared with the results of hydrodynamic calculations. It means that the contradiction in temperature reported~\cite{Swift-Seifter-Holtkamp-Clark-PRB-2007, Seifter-Swift-PRB-2008} is specific for the surface layer, but not the bulk.

For accurate calculations of the mentioned effects, one should exclude artificial viscosity or its analogs from the numerical scheme and use the physically based dissipative processes (the physical viscosity, the plasticity, and the heat conductivity) for stabilization of the solution instead. It is all the more actual since the calculations of the shock-wave processes in condensed matter are commonly performed on the fine numerical grids, which allow one to resolve the shock wave structure. We use the physical viscosity as a stabilizing factor, but the viscosity coefficient plays a role of parameter of numerical calculations here. Using a fine numerical grid allows one to obtain a stable solution at arbitrary small viscosity coefficient. Plasticity controls the shock wave thickness and the structure of the high- and low-entropy layers in the considered situations, those justify our approach. In spite of that, the determination of viscosity coefficients in condensed matter is necessary for a physically complete description of the problem.

\section*{Acknowledgements}
The authors greatly appreciate the useful discussions with A.A. Charakh\-ch'yan, I.R. Makeyeva, M.E. Povarnitsyn, V.F. Kuropatenko, D.A. Varfolomeyev, G.I. Kanel and A.Yu. Semenov.
This work is supported by grant No.\,18-08-01493 of the Russian Foundation for Basic Research as well as by act No.\,211 from 16 March 2013 of the Government of the Russian Federation (contract No.\,02.A03.21.0011) and by the Ministry of Science and Higher Education of the Russian Federation (state assignment for researches by JIHT RAS and CSU).





\hyphenation{Post-Script Sprin-ger Milyav-s-kii Khish-chen-ko Che-l-ya-b-insk
  But-ter-worth}

\end{document}